\documentclass[12pt,letterpaper]{article}
\usepackage{epsfig}
\usepackage{amsmath}
\usepackage{amssymb}
\usepackage{amsthm}
\usepackage{indentfirst}
\usepackage{xspace}
\usepackage{multirow}
\usepackage{hyperref}
\usepackage{xcolor}
\definecolor{darkred}{rgb}{0.5,0.15,0.15}
\hypersetup{colorlinks=true,urlcolor=darkred,linkcolor=darkred,citecolor=darkred}
\usepackage{verbatim}
\usepackage[letterpaper,margin=1in,headheight=15pt]{geometry}
\usepackage{mathpazo}
\usepackage{authblk}
\usepackage{empheq}

\usepackage[sorting=ydnt,bibstyle=authoryear-comp,labelyear=false,defernumbers=true,maxnames=20,firstinits=true, uniquename=init,dashed=false,backend=bibtex]{biblatex}

\DeclareNameAlias{sortname}{first-last}

\DeclareFieldFormat{url}{\url{#1}}
\DeclareFieldFormat[article]{pages}{#1}
\DeclareFieldFormat[inproceedings]{pages}{\lowercase{pp.}#1}
\DeclareFieldFormat[incollection]{pages}{\lowercase{pp.}#1}
\DeclareFieldFormat[article]{volume}{\textbf{#1}}
\DeclareFieldFormat[article]{number}{(#1)}
\DeclareFieldFormat[article]{title}{\MakeCapital{#1}}
\DeclareFieldFormat[inproceedings]{title}{#1}
\DeclareFieldFormat{shorthandwidth}{#1}

\DeclareBibliographyDriver{article}{%
  \usebibmacro{bibindex}%
  \usebibmacro{begentry}%
  \usebibmacro{author/editor}%
  \setunit{\labelnamepunct}\newblock
  \MakeSentenceCase{\usebibmacro{title}}%
  \newunit
  \printlist{language}%
  \newunit\newblock
  \usebibmacro{byauthor}%
  \newunit\newblock
  \usebibmacro{byeditor+others}%
  \newunit\newblock
  \printfield{version}%
  \newunit\newblock
  \usebibmacro{journal+issuetitle}%
  \newunit\newblock
  \printfield{note}%
  \setunit{\bibpagespunct}%
  \printfield{pages}
  \newunit\newblock
  \usebibmacro{eprint}
  \newunit\newblock
  \printfield{addendum}%
  \newunit\newblock
  \usebibmacro{pageref}%
  \usebibmacro{finentry}}

\renewbibmacro*{journal+issuetitle}{%
  \usebibmacro{journal}%
  \setunit*{\addspace}%
  \iffieldundef{series}
    {}
    {\newunit
     \printfield{series}%
     \setunit{\addspace}}%
  \printfield{volume}%
  \printfield{number}%
  \setunit{\addcomma\space}%
  \printfield{eid}%
  \setunit{\addspace}%
  \usebibmacro{issue+date}%
  \newunit\newblock
  \usebibmacro{issue}%
  \newunit}

\def\makebibcategory#1#2{\DeclareBibliographyCategory{#1}\defbibheading{#1}{\section*{#2}}}
\makebibcategory{books}{Books}
\makebibcategory{papers}{Refereed research papers}
\makebibcategory{chapters}{Book chapters}
\makebibcategory{conferences}{Papers in conference proceedings}
\makebibcategory{techreports}{Unpublished working papers}
\makebibcategory{bookreviews}{Book reviews}
\makebibcategory{editorials}{Editorials}
\makebibcategory{phd}{PhD thesis}
\makebibcategory{subpapers}{Submitted papers}
\makebibcategory{curpapers}{Current projects}

\renewcommand*{\bibitem}{\addtocounter{papers}{1}\item \mbox{}\hskip-0.85cm\hbox to 0.85cm{\hfill\arabic{papers}.~~}}
\defbibenvironment{bibliography}
{\list{}
  {\setlength{\leftmargin}{\bibhang}%
   \setlength{\itemsep}{\bibitemsep}%
   \setlength{\parsep}{\bibparsep}}}
{\endlist}
{\bibitem}

\newcounter{papers}
\newcounter{sumpapers}

\numberwithin{equation}{section}

\newcommand{\cL}{\ensuremath{\mathcal L}}

\newcommand{\cM}{\ensuremath{\mathcal M}}

\newcommand{\R}{\ensuremath{\mathbb R}}
\newcommand{\C}{\ensuremath{\mathbb C}}
\newcommand{\PP}{\ensuremath{\mathbb P}}
\newcommand{\Z}{\ensuremath{\mathbb Z}}

\newcommand{\ctM}{\widetilde{\cM}}
\newcommand{\half}{\ensuremath{\frac{1}{2}}}

\newcommand{\N}{{\mathcal N}}
\newcommand{\cA}{{\mathcal A}}
\newcommand{\F}{{\mathcal F}}

\newcommand{\hnabla}{{\widehat\nabla}}

\newcommand{\hk}{hyperk\"ahler\xspace}

\newcommand{\I}{{\mathrm i}}

\newcommand{\de}{\mathrm{d}}

\newcommand{\eps}{\epsilon}

\newcommand{\ti}[1]{\textit{#1}}

\DeclareMathOperator{\im}{Im}
\DeclareMathOperator{\re}{Re}

\bibliography{reduction-omega}

\begin{document}

\title{$\Omega$-deformed SYM on a Gibbons-Hawking Space}
\date{}
\author{Anindya Dey}
\affil{Theory Group and Texas Cosmology Center\\ University of Texas at Austin \footnote{PRE-PRINT NUMBER: UTTG-25-14}}
\maketitle

{\abstract{ We study an $\N=2$, pure $U(1)$ SYM theory on a Gibbons-Hawking space $\Omega$-deformed using the $U(1)$ isometry. The resultant 3D theory, after an appropriate ``Nekrasov-Witten" change of variables, is asymptotically equivalent to the undeformed theory at spatial infinity but differs from it as one approaches the NUT centers which are fixed points under the $U(1)$ action. The 3D theory may be recast in the form of a generalized hyperk\"ahler sigma model introduced in \cite{Dey:2014lja} where the target space is a one-parameter family of hyperk\"ahler spaces. The 
hyperk\"ahler fibers have a preferred complex structure which for the deformed theory depends on the parameter of $\Omega$-deformation. The metric on the hyperk\"ahler fiber can be reduced to a standard metric on $\C \times T^2$ with the modular parameter of the torus depending explicitly on the $\Omega$-deformation parameter. The contribution of the NUT center to the sigma model path integral, expected to be a holomorphic section of a holomorphic line bundle over the target space on grounds of supersymmetry, turns out to be a Jacobi theta function in terms of certain "deformed" variables.
}}

\pagebreak

\tableofcontents

\section{Introduction and Main Results}
\subsection{Basic Idea}
 Seiberg and Witten, in the seminal paper \cite{Seiberg:1996nz}, studied the IR Lagrangian for $\N=2$ supersymmetric theories in four dimensions compactified on $S^1 \times \R^3$, with $S^1$ of fixed radius $R$. Around a generic point of its moduli space, the IR Lagrangian is a  sigma model with a \hk target space $\cM[R]$ and a metric which depends in a non-trivial way on the radius $R$. In the context of wall-crossing phenomena for $\N=2$ supersymmetric theories, corrections to the \hk metric $g[R]$ due to massive BPS particles in the limit of large $R$ were studied in   \cite{Gaiotto:2008cd}.
   An interesting generalization of the Seiberg-Witten story involves studying the compactification of $\N=2$ theories on 4-manifolds where the circle direction is fibered non-trivially on the $\R^3$ base with isolated points where the fiber degenerates. Gibbons-Hawking space or multi-centered Taub-NUT space ($GH$) is a special example of such spaces which preserve 4 out of the 8 real supercharges of $\N=2$ supersymmetry algebra. The metric on $GH$ locally has the form
\begin{equation} \label{gh-metric}
\de s^2 = V(\vec{x}) \de{\vec{x}}^2 + \frac{R^2}{V(\vec{x})}(\de\chi - B)^2
\end{equation}
where $\vec{x}$ is a coordinate in $\R^3$, $V(\vec x)$ a harmonic function on $\R^3$ with isolated (coordinate) singularities 
and $B$ a 1-form on $\R^3$ which obeys $\star^{(3)} \de V = R \, \de B$. This is a \hk manifold with an $SU(2)$ holonomy rather than the generic $SU(2) \times SU(2) / \Z_2$, and this reduced holonomy admits 4 covariantly constant spinors. The preserved SUSY, however, is \emph{not} $\N=1$ in 4D, as should be obvious from the fact that the metric in \eqref{gh-metric} breaks translation symmetry.

The metric on $GH$ has an $SU(2) \times U(1)$ isometry, where the $U(1)$ vector field generates translations along the circle fiber. The coordinate singularities in $V(\vec x)$, also called NUT centers, are fixed points under this $U(1)$ isometry.\\

In \cite{Dey:2014lja}, the study of $\N=2$ theories on Gibbons-Hawking spaces was initiated and the effective low-energy description after reduction along the circle fiber was formulated.  The resultant 3d theory turns out to be a generalized version of the \hk sigma model and has two important ingredients which may be summarized as follows:
\begin{itemize}
\item {\bf Local Lagrangian:} For $\N=2$ theories compactified on $S^1 \times \R^3$, the low-energy effective theory can be formulated in terms of a local Lagrangian which is simply a \hk sigma model in 3d \cite{Bagger:1983tt}. Away from the NUT centers, the low-energy effective description of $\N=2$ theories on a Gibbons-Hawking space is also given by a local Lagrangian. It is a sigma model with a target space $\ctM$ -- a one-parameter \ti{family} of \hk spaces (of quaternionic dimension $r$) parametrized by the scalar $\varphi^0=\frac{V(\vec{x})}{R}$ -- which we treat as a $(4r + 1)$-real dimensional space. The sigma model Lagrangian involves a (possibly degenerate) bilinear form $g$ on $T \ctM$, which restricts on each constant $\varphi^0$ fiber to a \hk metric. Being \hk, the fibers of $\ctM$ carry a $\C\PP^1$ worth of complex structures.
While all these would be on the same footing in the usual \hk sigma model,  one of them is preferred in the deformed
model. 

Aside from the standard terms in the sigma model, the Lagrangian involves one extra
coupling, of the schematic form \begin{equation} \label{new-cs-term}
 \frac{1}{8\pi} \int_{\R^3} \de B \wedge \varphi^* A
\end{equation}
where $A$ represents a $U(1)$ connection in a line bundle $\cL$ over the family $\ctM$, and $\varphi^* A$ is its
pullback to $\R^3$ via the sigma model field $\varphi$. \\

\item {\bf NUT Operator:} Dimensional reduction procedure needs to be modified for the NUT centers where the circle fiber shrinks to zero. For every NUT center, we cut out a ball of radius $L$ and study the theory on a manifold with boundaries in a limit $E \ll 1/L \ll 1/R$ -- each boundary has the topology of $S^3$ in 4d and that of $S^2$ in 3d. The contribution of each such boundary to the sigma model action is a boundary operator $Q$, which 
in the limit $E \ll 1/L \ll 1/R$, is simply given by the lowest term in the derivative expansion -- an operator $Q(\phi)$ which is a function of constant boundary fields $\phi$ along $S^2$. Supersymmetry imposes strong constraints $Q(\phi)$. Defining $k = \frac{1}{4\pi} \int_{S^2} \de B$,  $Q(\varphi)$ obeys
\begin{equation} \label{pert-holomorphy}
 \bar\partial Q + k A^{(0,1)} = 0.
\end{equation}
where $\bar\partial$ and $A^{(0,1)}$ are defined with respect to the preferred complex structure on the fiber $\cM[\varphi^0]$. 

Geometrically, for \eqref{pert-holomorphy} to make global sense, $e^Q$ should be a \ti{holomorphic} section of $\cL^k$, where $\cL$ is the line
bundle introduced above with the $U(1)$ connection $A$ --with respect to a holomorphic structure on $\cL$ determined by $A^{(0,1)}$.
Since any two such sections will differ by a global holomorphic
function, in the preserved complex structure, and $\cM$ has rather few global holomorphic functions, \eqref{pert-holomorphy} is a very strong constraint on $Q$.\\
\end{itemize}

While the role of the preferred complex structure was clear in \cite{Dey:2014lja}, it is natural to ask whether other complex structures on the \hk fiber play any role in the generalized sigma model. The present work clarifies this specific issue -- it shows that other complex structures arise naturally if one considers an $\Omega$-deformed theory on a Gibbons-Hawking space (using the $U(1)$ isometry along the circle fiber) and makes certain change of variables (Nekrasov-Witten transformation) \cite{Nekrasov:2010ka} such that the deformed theory is  asymptotically equivalent to the undeformed one up to some trivial rescaling of parameters. In fact, the Nekrasov-Witten transformation ensures that the $\Omega$-deformed theory is equivalent to the undeformed theory everywhere that the $U(1)$ action is free. However, for a Gibbons-Hawking space the $U(1)$ action has a fixed point at the NUT center and therefore the theory is no longer equivalent to the undeformed one as one approaches the NUT center. We will demonstrate that the generalized sigma model associated to the $\Omega$-deformed theory, combined with this change of variables, has a preferred complex structure which is different from that of the undeformed theory and depends on the $\Omega$-deformation parameter. In particular, the NUT center operator in the deformed theory will be a holomorphic section with respect to this new complex structure. We mainly focus on the case of $\N=2, U(1)$ SYM and among other things, work out the exact NUT center operator in the deformed theory.

\subsection{Review of the generalized \hk sigma model}\label{genhk-rev}
Let us review the generalized \hk sigma model as introduced in \cite{Dey:2014lja}. It is convenient to first discuss the standard \hk sigma model \cite{Bagger:1983tt, Bagger:1982fn, AlvarezGaume:1981hm} and then extend it to the generalized case.

\subsubsection*{Standard \hk sigma model}  
Recall that the standard \hk sigma model describes the IR theory of an $\N=2$ SYM on $\R^3 \times S^1$. On the Coulomb branch, the gauge group $G$ is higgsed to $U(1)^{r}$ ($r=rank(G)$) by vevs of the adjoint scalars -- the IR theory therefore involves $4r$ scalars -- $3$ adjoint scalar and one dual photon for each $U(1)$ factor. The \hk sigma model is completely specified by the field content of the theory and certain differential geometric data on the target space $\cM$, as described below.\\

{\bf Field Content}
\begin{itemize}
\item The $4r$ scalar fields in the sigma model are $ \varphi^i \;: \R^3 \to \cM$, where $i=1,2,\ldots,4r$. $Sp(r)$ holonomy of the target space implies the following decomposition of the complexified tangent bundle $T_\C \cM = H \otimes E$, where $H$ is a complex rank $2$ trivial bundle with $SU(2)$ structure group and $E$ is a complex rank $2r$ bundle with structure group $Sp(r)$.  \footnote{ We will use unprimed uppercase Latin letters for $Sp(1)$ indices and primed uppercase Latin letters for $Sp(r)$ indices.}\\
\item The 4r 2-complex component fermions are $ \psi^{A'}_{\alpha},\bar{\psi}^{A'}_{\alpha}$ -- valued in the $Sp(r)$ bundle.\\
\item 8-complex dimensional  space of SUSY parameters $(\zeta^E_\alpha, \bar{\zeta}^E_\alpha)$ -- valued in the $SU(2)$ bundle.\\
\end{itemize}

{\bf Differential Geometric Data on $\cM$}
\begin{itemize}
\item A metric $g$ on $\cM$ and a Levi Civita connection -- written as $\Gamma^k_{ij}$ in local coordinates.\\
\item A connection in the $Sp(r)$ bundle -- $q^{A'}_{i B'}$ in local coordinates.\\
\item A "frame" on $\cM$, which is given by the isomorphism $e: T_\C \cM \to H \otimes E$ and represented by $e_{i\;EE'}$ in local coordinates.
\item Constraint from supersymmetry requires that the covariant derivative of $e_{i EE'}$ must vanish. In local coordinates,
 \begin{equation}
\partial_j e_{iEE'} - q_{jE'}^{A'} e_{iEA'} = \Gamma_{ji}^k e_{kEE'} 
\end{equation}
\end{itemize}

{\bf Local Lagrangian}
\begin{itemize}
\item Explicitly, the local Lagrangian for the sigma model is given as
\begin{equation}
\begin{split}
4 \pi \mathcal{L} & = \half \partial_\mu \varphi^i \partial^\mu \varphi^j g_{ij} - \I \bar\psi_{\alpha A'} \gamma^{\mu \alpha}_\beta (\partial_\mu \psi^{A'\beta} + q^{A'}_{B' i} \partial_\mu \varphi^i \psi^{B' \beta}) + \half \Omega_{A'B'C'D'} (\psi^{A'}_{\alpha} \gamma^{\mu\alpha}_\beta \bar\psi^{B'\beta}) (\psi^{C'}_\delta \gamma_{\mu \omega}^\delta \bar\psi^{D'\omega}).
\end{split} 
\end{equation}
where $\Omega_{A'B'C'D'}$ is the Riemann tensor contracted with $e^i_{AA'}$ and the antisymmetric $SU(2)$ tensor $\epsilon^{AB}$.\\
\item The SUSY transformation of the above action generated by 8-complex dimensional  space of SUSY parameters $(\zeta^E_\alpha, \bar{\zeta}^E_\alpha)$ is 
\begin{align}
\delta_\zeta \varphi^i &=  \psi^{E'\alpha} \bar\zeta^E_\alpha e^i_{EE'} + \bar\psi_{E'\alpha} \zeta^\alpha_E e^{i E E'}, \label{susyvar1}\\
\delta_\zeta \psi^{A'}_\alpha &= - \I \partial_\nu \varphi^i e_i^{EA'} \gamma_\alpha^{\nu \sigma} \zeta_{\sigma E} - q_{i E'}^{A'} \delta_\zeta \varphi^i \psi^{E'}_\alpha, \label{susyvar2}\\
\delta_\zeta \bar \psi^{\alpha}_{A'} &= -\I \partial_\nu \varphi^i e_{iEA'} \gamma_\sigma^{\nu \alpha} \bar\zeta^{\sigma E} + q_{i A'}^{E'} \delta_\zeta \varphi^i \bar\psi_{E'}^\alpha. 
\end{align}

\end{itemize}

Having familiarized ourselves with the standard story of a \hk sigma model, one can now readily extend it to the generalized sigma model.
\subsubsection*{Generalized \hk sigma model} 
Since the generalized \hk sigma model gives the IR theory of an $\N=2$ SYM compactified on a Gibbons-Hawking space, its scalar fields must include $4r$ scalars present in the standard sigma model. In addition, there is a non-dynamical, background scalar $\varphi^0$ which is a harmonic function on $\R^3$. Roughly speaking, $\varphi^0$ controls the size of the circle fiber as one moves in $\R^3$.

The target space of the sigma model is a 1-parameter family of \hk manifolds parametrized by $\varphi^0$ --we refer to the total space as $\ctM$. As before, the sigma model is completely specified by the field content of the theory and certain differential geometric data on the target space $\ctM$, as described below.\\

{\bf Field Content}
\begin{itemize}
\item The $4r+1$ scalar fields in the sigma model are $ \varphi^i \;: \R^3 \to \ctM$, where $i=0,1,2,\ldots,4r$. Of these, $\varphi^0$ is a background, harmonic function on $\R^3$ and has a dual 1-form $B$ such that $\star \de B= \de \varphi^0$. The $SU(2)$ bundle $H$ and the $Sp(r)$ bundle $E$ are extended over the full space $\ctM$.
\item The 4r 2-complex component fermions are $ \psi^{A'}_{\alpha},\bar{\psi}^{A'}_{\alpha}$ -- valued in the $Sp(r)$ bundle.\\
\item 4-complex dimensional  space of SUSY parameters is given by $(\zeta^E_\alpha, \bar{\zeta}^E_\alpha)$ -- valued in the $SU(2)$ bundle-- subject to the following constraints
\begin{equation}
 c^E \zeta^\alpha_E = 0, \qquad c_E \bar\zeta^E_\alpha = 0.
\end{equation}
which reduce the number of preserved supersymmetry to 4 from 8. The supersymmetry parameters may depend explicitly on $\varphi^0$, such that
\begin{equation}
\begin{split}
&\partial_{\mu} \zeta^{\alpha}_E + f(\varphi^0) \partial_{\mu} \varphi^0\zeta^{\alpha}_E=0, \\
&\partial_{\mu} \bar{\zeta}^{\alpha}_E + f(\varphi^0) \partial_{\mu} \varphi^0\bar{\zeta}^{\alpha}_E=0.
\end{split} 
\end{equation}
\end{itemize}

{\bf Differential Geometric Data on $\ctM$}
\begin{itemize}
\item A degenerate metric $g$ on $\ctM$ which restricts to a \hk metric on each fiber $\ctM[\varphi^0]$.  The Levi Civita connection is naively extended to the full space $\ctM$ -- we write it as $\Gamma^k_{ij}$ (with $i,j,k=0,1,\ldots,4r$) in local coordinates. Note that one cannot simply derive this connection from the metric $g$ and to compute these one needs some prescription, which we state below. 
\item A connection in the $Sp(r)$ bundle now extended over $\ctM$ -- $q^{A'}_{i B'}$ in local coordinates. It is convenient to define a shifted version of the connection $\tilde{q}_{jE'}^{A'} = q_{j E'}^{A'} + f(\varphi^0) \delta^0_j \delta_{E'}^{A'}$, where $f(\varphi^0)$ is the function defined above.
\item A "frame" on $\ctM$, which is given by the surjection $e: T_\C \cM \to H \otimes E$ and represented by $e_{i\;EE'}$ in local coordinates.
\item A necessary condition for  supersymmetry is a certain generalization the covariant constancy of $e_{i EE'}$. In local coordinates,
 \begin{equation}
\partial_j e_{iEE'} - \tilde{q}_{jE'}^{A'} e_{iEA'} = \Gamma_{ji}^k e_{kEE'} \label{ext-hk-id1}
\end{equation}
Given $e_{i EE'}, q_{j E'}^{A'}$ and $f(\varphi^0)$, the above equation gives a prescription for computing $\Gamma_{ji}^k$.
\item A 1-form $\cA$ on $\ctM$ with curvature $\F$ -- $\F$ is a $(1,1)$ form with respect to the preferred complex structure on the \hk fiber.\\
\end{itemize}

{\bf The Local Lagrangian}  
\begin{itemize}
\item Explicitly, the local Lagrangian is given as ($i,j=0,1,\ldots,4r$)
\begin{equation}
\begin{split}
4 \pi \mathcal{L}_{loc} & = \half \partial_\mu \varphi^i \partial^\mu \varphi^j g_{ij} 
- \I \bar\psi_{\alpha A'} \gamma^{\mu \alpha}_\beta (\partial_\mu \psi^{A'\beta} + q^{A'}_{B' i} \partial_\mu \varphi^i \psi^{B' \beta}) \\
&+ \half \Omega_{A'B'C'D'} (\psi^{A'}_{\alpha} \gamma^{\mu\alpha}_\beta \bar\psi^{B'\beta}) (\psi^{C'}_\delta \gamma_{\mu \omega}^\delta \bar\psi^{D'\omega}) 
+ \half \eps_{\mu \nu \rho} G^{\mu\nu} \partial^\rho \varphi^i \cA_i.
\end{split} \label{genhksigma}
\end{equation}
Note that in the special case where $G = 0$ and $\varphi^0$ is constant, this action reduces
to the undeformed \hk sigma model as written above. 
    The form of the SUSY transformations generated by fermionic parameters $\zeta^\alpha_E, \bar\zeta^E_\alpha$, which now obey constraint equations, remain the same and $\delta \varphi^0=0$, since $\varphi^0$ is a background field.\\

\item  The sufficient condition for the above action to preserve 4 supercharges \cite{Dey:2014lja} is
\begin{equation}
\begin{split}
&c^E e^l_{EE'} \left( \partial_i g_{jl} + \partial_j g_{il} - \partial_l g_{ij} - ({\Gamma}_{ij}^k+{\Gamma}_{ji}^k) g_{kl} + \delta_{j}^0 \F_{il} + \delta_i^0 \F_{jl} \right) = 0,  \\
&c^E e_{kEE'} (\Gamma^k_{ji} - \Gamma^k_{ij}) = 0,
\end{split} \label{susyloc}
\end{equation}
where the preserved supercharges correspond $c^E = (0,1)$. \\
It is convenient to define a related connection $\hnabla$ on $\ctM$ in the following fashion: let $\hat{\Gamma}^k_{ij}$ agree with $\Gamma^k_{ij}$ when $k$ is holomorphic and $\hat{\Gamma}^k_{ij}$ for $k$ anti-holomorphic are determined by the reality of $\hnabla$. Then the second equation in $\eqref{susyloc}$ simply implies that the connection $\hnabla$ is torsionless. The first equation for $i,j,l \neq 0$ simply implies that the connection restricts fiberwise to a a Levi-Civita connection, as expected. The non-trivial equation arises from $i=0$ with $j,l$ arbitrary, namely
\begin{equation}
\boxed{\hnabla_0 g_{ij} + \F_{ij}=0}
\end{equation}
The metric and the curvature $\F$ (including all quantum corrections in a generic theory) should obey the above equation purely on grounds on supersymmetry.

\end{itemize}

{\bf The NUT Operator}
\begin{itemize}
\item Including the contribution of the NUT center, the action of the generalized sigma model is 
\begin{equation}
S=\int _{X} \mathcal{L}_{loc} + \sum_i Q_i(\varphi(0)) \label{sigma-action-full}
\end{equation}
where $X$ is the manifold described earlier with an $S^2$ boundary for each NUT center and  $Q_i(\varphi(0))$ is a function of constant field configurations along the $i$-th $S^2$.
\item Supersymmetry imposes a very strong constraint on the operator $Q$, namely
\begin{equation}
c^E e^i_{EE'} (\partial_i + k A_i)e^Q =0 \label{susynut}
\end{equation}
which implies that $e^Q$ is a holomorphic section of a holomorphic line bundle described above.
\item In the case of a $U(1)$ SYM on NUT space, the section $e^Q$ can be exactly computed and is found to be related to the Jacobi theta function.
\begin{equation}
e^Q = \Psi\left(\theta_e,\theta_m,\tau \right) = e^{\frac{\I}{2\pi}(\tau \theta_e^2/2-\theta_e \theta_m)} \Theta (\tau,2y )
\end{equation}
where $y=\frac{\theta_m-\tau \theta_e}{4\pi}$, $\bar{y}=\frac{\theta_m-\bar{\tau} \theta_e}{4\pi}$; $\theta_e$ is the asymptotic holonomy of the $U(1)$ gauge field, $\theta_m$ is a periodic scalar obtained on dualizing the photon and $\tau$ is the complexified gauge coupling.
\end{itemize}

\subsection{Main results of this paper}
\begin{itemize}
\item Starting from the standard 6D description of an $\N=2, U(1)$ pure SYM on a Gibbons-Hawking space $\Omega$-deformed using the $U(1)$ isometry (which corresponds to translations along the circle fiber), we explicitly derive the associated 3D generalized \hk sigma model via dimensional reduction. For simplicity, we take the $\Omega$-deformed space to be a Gibbons-Hawking bundle over $S^1$ as opposed to the full torus, taking the $\Omega$-deformation parameter $\epsilon$ to be real. The resultant 6D metric on the $\Omega$-deformed $GH \times T^2$ space is 
\begin{equation}
ds^2_6=\sum^3_{m=0} (e^m_{(GH)} -\epsilon V^m \de u )^2 + (\de u)^2 - (\de v)^2 
\end{equation}
where $\{e^m_{(GH)}\}$ are vierbeins on Gibbons-Hawking space and $V^m$ is the $U(1)$ Killing vector field used to implement the $\Omega$-deformation.
After performing a ``Nekrasov-Witten"-like change of variables at the level of the 3D theory, one can again recast it as a generalized \hk sigma model which is asymptotically equivalent to the undeformed sigma model at spatial infinity up to certain rescaling of the radius $R$ and the gauge coupling $\tau$.
\begin{equation}
\begin{split}
&R'=\frac{R}{\sqrt{1+\epsilon^2 R^2 }},\\
&\re \hat{\tau}= \re \tau,\\
&\im \hat{\tau}= \im \tau \sqrt{1+\epsilon^2 R^2 } .
\end{split} \label{rescaled-par}
\end{equation}
However, the theory differs from the undeformed sigma model as one approaches fixed points under the $U(1)$ action -- the NUT centers. 

\item Our computation of the local action for the deformed theory allows one to simply read off the relevant sigma model data -- the metric $g$, the 1-form $\cA$, $e_{iEE'}$ and the $Sp(r)$ connection -- as explicit functions of the $\Omega$-deformation parameter. Supersymmetry constraint for the local action takes the same form as before \eqref{susyloc} with a deformed $\F$, $g$ and $c^E$.
\item The target space of the generalized \hk sigma model $\ctM$ is a $(4+1)$-real dimensional space which for constant $\varphi^0(=\frac{V(\vec{x})}{R'})$ restricts to a \hk fiber of quaternionic dimension $1$. As discussed earlier, there exists a preferred complex structure parametrized by the $Sp(1)$ vector $c^E$ on the \hk fiber. In the undeformed theory, $c^E$ had the simple form: $c^E=(0,1)$. The deformed sigma model obtained by dimensional reduction of the $\Omega$-deformed theory, combined with the said change of variables, has a preferred complex structure  which depends on the $\Omega$-deformation parameter $\epsilon$. This is reflected in the appearance of  a rotated version of the $Sp(1)$ vector in the deformed sigma model, namely 
\begin{equation}
c^E=  (-\sin{\frac{\theta}{2}}, \cos{\frac{\theta}{2}}). \label{def-c}
\end{equation}
The angle $\theta$ is related to the parameter of $\Omega$-deformation $\epsilon$ by the equation:
\begin{equation}
\boxed{\cos{\theta} =\frac{1}{\sqrt{1+\epsilon^2 R^2 }}}
\end{equation}
Recall that the curvature $\F$ of the 1-form $\cA$ in the undeformed theory was $(1,1)$ in the preferred complex structure of that theory. Similarly, the curvature $\F$ in the deformed theory is a $(1,1)$ in the new preferred complex structure.\\

We would like to emphasize that the $Sp(1)$ vector $c^E$ obtained in \eqref{def-c} does not parametrize the most generic complex structure. Since our $\Omega$-deformation parameter $\epsilon$ is real, we are restricted to a 1-parameter subset of all possible complex structures that may arise in this fashion. 

\item The metric on a constant $\varphi^0$ slice of the target space in the deformed theory can be reduced to a standard metric on $\C \times T^2$. The modular parameter $\tilde{\tau}$ of the torus now explicitly depends on the $\Omega$-deformation parameter and $V(\vec{x})$.
\begin{equation}
\begin{split}
& \re \tilde{\tau} =\re {\tau},\\
& \im \tilde{\tau} =\im {\tau} \sqrt{\frac{(V(\vec{x})\cos^2{\theta}+\sin^2{\theta})}{V(\vec{x}) \cos^2{\theta}}}.
\end{split}
\end{equation}

\item As guessed in \cite{Neitzke:2011za}, the deformed NUT operator $e^{\widetilde{Q}}$ turns out to be a holomorphic section with respect to the new preferred complex structure. The supersymmetry constraint is of the same form as in \eqref{susynut}, i.e.
\begin{equation}
\boxed{c^E e^i_{EE'} (\partial_i + k A_i)e^{\widetilde{Q}} =0,} \label{NUT-def-susy}
\end{equation}
where $c^E$ is now given by \eqref{def-c} and $e_{iEE'}$, corresponding to the deformed theory, are given in \eqref{e-def-2}.

\item Finally, we compute the deformed NUT operator for the $U(1)$ theory and the answer expectedly turns out to be a Jacobi theta function in some ``deformed" variable $z$ which is a combination of various boundary values of fields at spatial infinity -- $\theta_e$ which is the asymptotic holonomy of the gauge field, $\theta_m$ which is associated with the dual photon and $u$ which is the asymptotic value of the scalar $A_u$ at spatial infinity. In addition, the operator depends on the $\Omega$-deformation parameter. Explicitly,
\begin{empheq}[box=\fbox]{gather}
e^{\widetilde{Q}} = \Psi\left(\theta_e,\theta_m, u,\hat{\tau}, \theta \right)
= e^{\frac{\I}{2\pi}(\hat{\tau} \widetilde{\theta}_e^2/2-\widetilde{\theta}_e \widetilde{\theta}_m)} \Theta (\hat{\tau},2z). \label{Nut-op-intro}
\end{empheq}
The "deformed"variables in the above equation are: 
\begin{align}
&\widetilde{\theta}_m=\theta_m +4\pi R' u \re \tau \tan{\theta},\\
&\widetilde{\theta}_e=\theta_e +4\pi R' u \re \tau \tan{\theta},\\ 
&z = \widetilde{\theta}_m -\hat{\tau} \widetilde{\theta}_e,
\end{align}
where $R'$ is the rescaled radius  and $\hat{\tau}$ is the rescaled coupling constant. The supersymmetry constraint \eqref{NUT-def-susy}, in terms of the local coordinates $z,\bar{z}$ defined above reduces to the following simple form:
\begin{empheq}[box=\fbox]{gather}
\left(\partial_{\bar{z}} - \mathcal{A}_{\bar{z}}\right) e^{\widetilde{Q}} =0.
\end{empheq}
It is straightforward to check that the operator $e^{\widetilde{Q}}$ in \eqref{Nut-op-intro} indeed satisfies the constraint.\\
\end{itemize}
The rest of the paper is organized in the following fashion. Section 2 describes the $\Omega$-deformation of an $\N=2, U(1)$ gauge theory on a Gibbons-Hawking space using the $U(1)$ isometry along the circle fiber and derives the associated generalized \hk sigma model obtained by simply dimensionally reducing the 4d action. Section 3 discusses the Nekrasov-Witten change of variables in an $\Omega$-deformed theory on $\R^3 \times S^1$ from a  3d point of view. Section 4 discusses the Nekrasov-Witten change of variables in the $\Omega$-deformed theory on a Gibbons-Hawking space and the corresponding sigma model picture. Finally, section 5 derives the NUT operator in the $\Omega$-deformed theory after Nekrasov-Witten change of variables.

\section*{Acknowledgements}
The author would like to thank Andrew Neitzke for numerous discussions on this project and related topics. The author would also like to thank Mina Aganagic, Nikita Nekrasov and Martin Ro\v{c}ek for discussion and useful comments.\\

This material is based upon work supported by the National Science Foundation under Grant Number PHY-1316033.

\section{$\Omega$-deformed $\mathcal{N}=2$ $U(1)$ theory on Gibbons-Hawking Spaces}
In this section, we derive the 4d action of an  $\Omega$-deformed $\N=2, U(1)$ SYM on a Gibbons-Hawking space starting from an $\N=1, U(1)$ SYM in 6d, dimensionally reduce to 3d and write down the corresponding generalized \hk sigma model. As mentioned earlier, we use the Killing vector for the $U(1)$ isometry of a Gibbons-Hawking space to $\Omega$-deform the 6d background.
\subsection{6d Description}
Consider the undeformed 6d background ${GH} \times T^2$ where ${GH}$ is a Gibbons-Hawking space. $\Omega$-deformation  from a six-dimensional perspective is simply deforming the metric in the following fashion:
\begin{equation}
ds^2_6=\sum^3_{m=0} (e^m_{(GH)} -\epsilon V^m \de u )^2 + (\de u)^2 - (\de v)^2  \label{6d.omega-manifold}
\end{equation}
where $m=0,1,2,3$ denote the Lorentz indices on ${GH}$ and $V=\frac{\partial}{\partial \chi}$ is the Killing vector field which generates  the $U(1)$ isometry corresponding to translations along the circle direction. In the Lorentz basis, the components of the vector field are $V^i=0 (i=0,1,2),\; V^3=\frac{R}{\sqrt{V}}$.\\
To illustrate the deformation more explicitly, consider the example of a single-centered Taub-NUT space. In terms of spherical polar coordinates in 4d, the deformed metric can be written as
\begin{equation}
ds^2_6=V(r) (\de r^2 +r^2 \de \Omega^2_2 ) +\frac{1}{V(r)}(\de y + R\cos{\theta}\de \phi -\epsilon R \de u)^2 + (\de u)^2 - (\de v)^2 
\end{equation}
where $V=1+\frac{R}{r}$ and we define the coordinate $y$ as $dy = Rd\chi$, such that it is periodic : $y \sim y + 4\pi R$.\\
One can now choose the following orthonormal basis of vector fields:
 \begin{equation}
e^M : e^i =e^i_{TN}, \; e^3 =e^3_{TN} -\frac{\epsilon R}{\sqrt{V}} \de u, \; e^4=\de u,\; e^5= \de v
\end{equation}
In matrix notation ($\mu$ labels rows and $M$ labels columns):
\begin{equation}
e^{\;\;M}_{\mu} =\left(
\begin{array}{cccccc}
 \sqrt{V[r]} & 0 & 0 & 0 & 0 & 0 \\
 0 & r \sqrt{V[r]} & 0 & 0 & 0 & 0 \\
 0 & 0 & r \text{Sin}[\theta ] \sqrt{V[r]} & \frac{R \text{Cos}[\theta ]}{\sqrt{V[r]}} & 0 & 0 \\
 0 & 0 & 0 & \frac{1}{\sqrt{V[r]}} & 0 & 0 \\
 0 & 0 & 0 & -\frac{R \epsilon }{\sqrt{V[r]}} & 1 & 0 \\
 0 & 0 & 0 & 0 & 0 & 1
\end{array}
\right)
\end{equation}
while the inverse (with $M$ labeling rows and $\mu$ labeling columns) is given as
\begin{equation}
E_{M}^{\;\;\;\; \mu} =\left(
\begin{array}{cccccc}
 \frac{1}{\sqrt{V[r]}} & 0 & 0 & 0 & 0 & 0 \\
 0 & \frac{1}{r \sqrt{V[r]}} & 0 & 0 & 0 & 0 \\
 0 & 0 & \frac{\text{Csc}[\theta ]}{r \sqrt{V[r]}} & -\frac{R \text{Cot}[\theta ]}{r \sqrt{V[r]}} & 0 & 0 \\
 0 & 0 & 0 & \sqrt{V[r]} & 0 & 0 \\
 0 & 0 & 0 & R \epsilon  & 1 & 0 \\
 0 & 0 & 0 & 0 & 0 & 1
\end{array}
\right)
\end{equation}

The generalization of the above to a generic Gibbons-Hawking case is straightforward. Now, we put a pure $\N=1, U(1)$ SYM on the deformed 6-manifold and dimensionally reduce to 4d along the torus to obtain the deformed $\N=2$ theory on ${GH}$.\\

The 6d Lagrangian for $\N=1, U(1)$ SYM can be written as ($L^2=$Area of the torus)
\begin{equation}
 \begin{split}
 S=& \frac{1}{g^2_{YM} L^2} \; \int d^6 x \sqrt{g_{\Omega}} \left[ \mathcal{L}_b +\mathcal{L}_f \right]\\
 \mathcal{L}_b=& \frac{1}{2} F_{MN} F_{MN} = \frac{1}{2} F_{mn} F_{mn} +\left( F_{m4}^2 - F^2_{m5} - F^2_{45}\right)=:\mathcal{L}^{(1)}_b + \mathcal{L}^{(2)}_b\\
 \mathcal{L}_f =&\bar{\psi}_a \Gamma_M D_M \psi^a=\bar{\psi}_a \Gamma_m D_m \psi^a + \left(\bar{\psi}_a \Gamma_4 D_4 \psi^a -\bar{\psi}_a \Gamma_5 D_5 \psi^a \right)=: \mathcal{L}^{(1)}_f + \mathcal{L}^{(2)}_f
 \end{split}
\end{equation}
where $g_{\Omega}$ is the metric on the $\Omega$-deformed space. Here, we have separated the bosonic and fermionic terms into two groups for convenience.

The rules of supersymmetry for the fields are
\begin{equation}
\begin{split}
&\delta A_M = -\bar{\psi}_a \Gamma_M  \zeta^a\\
&\delta \psi^a=  -\frac{1}{2} F_{MN} \Gamma_{MN} \zeta^a\\
&\delta \bar{\psi}^a=  \frac{1}{2} F_{MN} \bar{\zeta}^a\Gamma_{MN} 
\end{split}
\end{equation}
Here, $\zeta^a$ is a covariantly constant spinor on the manifold \eqref{6d.omega-manifold}. For the case of NUT space, this spinor is explicitly derived in Appendix \S\ref{KS-omega} -- generalization to arbitrary Gibbons-Hawking space is straightforward.

To implement the dimensional reduction of the above action to 4d, it is instructive to rewrite the three terms labeled $\mathcal{L}^{(2)}_b$ in some detail. Note that
\begin{equation}
\begin{split}
&D_m A_4 = E^{\mu}_{m} \left( D_{\mu} A_u + \epsilon R D_{\mu} A_y \right)=  E^{\mu}_{m} \left( \partial_{\mu} A_u + \epsilon R \partial_{\mu} A_y \right)\\
&D_4 A_m=  E^{\mu}_{m} \left(D_{u} A_{\mu}+\epsilon R D_{y} A_{\mu} \right)= E^{\mu}_{m} \left(\partial_{u} A_{\mu}+\epsilon R \partial_{y} A_{\mu} \right)\\
&F_{m4}=E^{\mu}_m \left( F_{\mu u}  + \epsilon R F_{\mu y}\right)=F_{mu} + \epsilon R F_{m y}\\
&F_{m5}= E^{\mu}_{m} \left( \partial_{\mu} A_{v} - \partial_{v} A_{\mu}\right)=F_{mv}\\
&F_{45}= \left(F_{uv} + \epsilon R  F_{yv} \right)
\end{split}
\end{equation}

The 6D action can then be written as
\begin{equation}
 \begin{split}
 S=& \frac{1}{g^2_{YM} L^2} \; \int d^6 x \sqrt{g_{\Omega}} \left[ \mathcal{L}_b +\mathcal{L}_f \right]\\
 \mathcal{L}_b=& \frac{1}{2} F_{mn} F_{mn} +F^2_{3u}+\sum_i\left( F_{iu}+\epsilon R F_{iy}\right)^2 - \sum_m F^2_{mv} -\left(F_{uv} + \epsilon R  F_{yv} \right)^2 \\
 =& \frac{1}{2} F_{mn} F_{mn}+\sum_m F^2_{mu}-\sum_m F^2_{mv} - F^2_{uv} \\
 & + \sum_i \left(2 \epsilon R F_{iu} F_{iy}+\epsilon^2 R^2 F^2_{iy} \right) - \left(2 \epsilon R F_{uv} F_{yv}+\epsilon^2 R^2 F^2_{yv} \right)\\
 \mathcal{L}_f =&\bar{\psi}_a \Gamma_m D_m \psi^a + \bar{\psi}_a \Gamma_4 D_u \psi^a +\bar{\psi}_a \Gamma_5 D_v \psi^a +\epsilon R \bar{\psi}_a \Gamma_4 D_y \psi^a
 \end{split}
\end{equation}

Note that the operator $D_v$ simply acts as $\partial_v$ while the operator $D_u+\epsilon R D_y $ acts as $\partial_u + \epsilon R \partial_y$ as shown in Appendix \S\ref{KS-omega} -- spin connection terms in the differential operator $D_u+\epsilon R D_y $ simply cancel out.

The rules of SUSY transformation can also be rewritten as
\begin{equation}
\begin{split}
\delta A_i = &-\bar{\psi}_a \Gamma_i  \zeta^a, \; \delta A_y = -\bar{\psi}_a \Gamma_3  \zeta^a/\sqrt{V}\\
\delta A_u = &-\bar{\psi}_a \Gamma_4  \zeta^a + \frac{\epsilon R}{\sqrt{V}} \bar{\psi}_a \Gamma_3  \zeta^a,\; \delta A_v = -\bar{\psi}_a \Gamma_5  \zeta^a\\
\delta \psi^a= & -\frac{1}{2} F_{mn} \Gamma_{mn} \zeta^a- F_{3u} \Gamma_{34} \zeta^a- (F_{iu}+ \epsilon R F_{iy})\Gamma_{i4} \zeta^a+F_{mv} \Gamma_{m5} \zeta^a+(F_{uv}+\epsilon R F_{yv} ) \Gamma_{45} \zeta^a\\
\delta \bar{\psi}^a= & -\frac{1}{2} F_{mn}\bar{\zeta}^a \Gamma_{mn} - F_{3u} \bar{\zeta}^a\Gamma_{34} - (F_{iu}+ \epsilon R F_{iy})\bar{\zeta}^a\Gamma_{i4} +F_{mv} \bar{\zeta}^a\Gamma_{m5} +(F_{uv}+\epsilon R F_{yv} ) \bar{\zeta}^a\Gamma_{45} 
\end{split} 
\end{equation}

\subsection{Dimensional Reduction to Four Dimensions}
Killing vectors along the three circle directions of the $\Omega$-deformed 6-manifold are :
\begin{equation}
\begin{split}
V=&\frac{\partial}{\partial v}\\
U=&\frac{\partial}{\partial u} + \epsilon R \frac{\partial}{\partial y}\\
Y=&  \frac{\partial}{\partial y}
\end{split}
\end{equation}
Dimensional reduction of the theory to 4D is implemented by imposing that the Lie derivatives of the fields w.r.t the Killing vectors $V$ and $U$ vanish.
\begin{equation}
\mathcal{L}_X A_{\mu} =0, \mathcal{L}_X \psi^a=0 \;\; \mbox{with } X=V,U.
\end{equation}

The 4D action can then be written as
\begin{equation}
 \begin{split}
 S=& \frac{1}{g^2_{YM}} \; \int d^4 x \sqrt{g_{TN}} \left[ \mathcal{L}_b +\mathcal{L}_f \right]\\
 \mathcal{L}_b=& \frac{1}{2} F_{mn} F_{mn} +(E^y_3)^2(\partial_y A_u+ \epsilon R \partial_y A_y)^2+\sum_i(E^\mu_i)^2\left( \partial_{\mu} A_u+\epsilon R \partial_{\mu} A_y\right)^2 - \sum_m (E^\mu_m)^2(\partial_\mu A_v)^2 \\
 \mathcal{L}_f =&-2\I \bar{\lambda}_a \bar{\sigma}^m D_m \lambda^a
 \end{split}
\end{equation}

The SUSY transformation, generated by the chiral half of the supersymmetry parameters on $\mathbb{R}^3 \times S^1$, may be summarized as: 
\begin{equation}
\begin{split}
&\delta A_i= -\I\zeta_{a,{\alpha}} \sigma^{\alpha \dot{\beta}}_i \bar{\lambda}^a_{\dot{\beta}}\; (i=0,1,2),\\
& \delta A_y=-\I\zeta_{a,{\alpha}} \sigma^{\alpha \dot{\beta}}_3 \bar{\lambda}^a_{\dot{\beta}}/\sqrt{V},\\
& \delta A_u =- \zeta^{\alpha}_{a}\lambda_{\alpha}^{a}+\I\frac{\epsilon R}{\sqrt{V}}\zeta_{a,\alpha} \sigma^{\alpha \dot{\beta}}_3 \bar{\lambda}^a_{\dot{\beta}} ,\\
&\delta A_v =-\zeta^{\alpha}_{a}\lambda_{\alpha}^{a},\\
&\delta \lambda^{a,\alpha} = \frac{1}{2} F_{mn}(\sigma^{mn})^{\alpha}_{\;\;\beta}\zeta^{\beta, a}, \\
& \delta \bar{\lambda}^a_{\dot{\alpha}} = - \I{\zeta}^{a,\;{\alpha}} (\bar{\sigma}^{m})_{\dot{\alpha} {\alpha}} (\partial_m A_{u}+\epsilon R \partial_m A_{y})+\I{\zeta}^{a,\;{\alpha}} (\bar{\sigma}^{m})_{\dot{\alpha} {\alpha}} \partial_m A_{v}.
\end{split} \label{susy4D-4DspinorsNUT}
\end{equation}

One can now dimensionally reduce along the circle direction of the Gibbons-Hawking space to obtain the $\Omega$-deformed theory in 3D.

\subsection{Dimensional Reduction to Three Dimensions}

Dimensional reduction to three dimensions is implemented by imposing the following condition on the fields,
\begin{equation}
\mathcal{L}_Y A_{\mu} =0, \mathcal{L}_Y \lambda^a=0 
\end{equation}

 To implement the $4D \to 3D$ reduction, we decompose the four dimensional gauge field in the following basis of 1-forms: $A^{(4)}=A_\mu dx^\mu - R \sigma \Theta $, where $\Theta=d\chi +B$. Note that the field $\sigma$ is defined such that $\sigma=\frac{A_3}{\sqrt{V}}=-A_y$.
 
 The $\Omega$-deformed action reduced to 3D therefore reads:
\begin{equation}
S= \frac{R}{g^2_{YM}} \; \int_{\mathbb{R}^3} d^3 x  \left[ \mathcal{L}_b +\mathcal{L}_f \right]
\end{equation}
where the bosonic and the fermionic parts of the Lagrangian density are
\begin{equation}\label{3Dactionb4dual}
 \begin{split}
&\mathcal{L}_b= \frac{1}{2}V^{-1}(F^{(3)}_{\mu \nu})^2 + V (\partial_\mu \sigma)^2 +  \left(\partial_\mu A_u -\epsilon R\partial_\mu \sigma \right)^2 -  \left( \partial_\mu A_v \right)^2,\\
&\mathcal{L}_f= 2\I \sqrt{V} \bar{\lambda}_a \gamma^\mu \partial_\mu \lambda^a.
\end{split}
\end{equation}
where $F^{(3)}$ is defined as $F^{(3)} =dA^{(3)}-R \sigma dB$. \\

Therefore, in 3D, the action for the $\Omega$-deformed theory is formally obtained by $A_u \to A_u - \epsilon R \sigma$. It is important to emphasize that this deformation is {\it not} a change of variable -- since one is shifting a non-periodic scalar by a periodic scalar -- and indeed gives rise to a physically inequivalent theory.

The SUSY transformation rules can be summarized as
\begin{equation}
\begin{split}
& \delta A_\mu= -\I\sqrt{V}\zeta_{a} \gamma_\mu \bar{\lambda}^a\; (\mu=1,2,3),\\
& \delta \sigma=\frac{1}{\sqrt{V}}\zeta_a \bar{\lambda}^a,\\
& \delta A_u =-\zeta_{a}\lambda^{a}+\frac{\epsilon R}{\sqrt{V}}\zeta_a \bar{\lambda}^a,\\
&\delta A_v =-\zeta_{a}\lambda^{a}, \\
&\delta \lambda_{\alpha}^{a} =-\frac{\I}{2} V^{-1} \;F^{(3)}_{\mu \nu}\epsilon^{\mu \nu \kappa}(\gamma_{\kappa})_{\alpha}^{\;\;\beta}\zeta_{\beta}^{a} +\I \partial_\mu\sigma (\gamma^\mu)_{\alpha}^{\;\;\beta}\zeta_{\beta}^{a},  \\
&\delta \bar{\lambda}^a_{\alpha} = \frac{\I}{\sqrt{V}}(\gamma^{\mu})_{\alpha}^{\;\;\beta} {\zeta}^{a}_{\beta} \partial_\mu( A_{u}- A_{v} -\epsilon R \sigma ).
\end{split} \label{susy3D-3DspinorsNUTomega}
\end{equation}

Now consider dualizing the action which would allow one to recast the theory as a sigma model.  The details of the dualization procedure for an $\N=2, U(1)$ theory on  a Gibbons-Hawking space is described in Appendix \S\ref{dualization}.  Dualization introduces a second periodic scalar in the theory--the dual photon $\gamma$. Since $\Omega$-deformation only affects the adjoint scalar $A_u$ and not the gauge field, the dualized action for the deformed theory may be obtained from the undeformed one by a formal substitution: $A_u \to A_u - \epsilon R \sigma$. Therefore, the dualized action for the $\Omega$-deformed theory is 
\begin{equation}
\begin{split}
{S}^b_{(3)} &= R \int_{\mathbb{R}^3} \de^3 x \Big[V (\im \tau)^{-1} |\partial_\mu \gamma - \tau \partial_\mu  \sigma|^2+ \im \tau (\partial_\mu A_u - \epsilon R \partial_\mu \sigma)^2 
- \im \tau (\partial_\mu A_v)^2\Big]\\
&\;\;\;  -2\I R^2\int_{\mathbb{R}^3} \gamma \wedge \de\sigma \wedge dB.\\
{S}^f_{(3)} &= 2\I R\; \im \tau\int_{\mathbb{R}^3} \de^3x\;  \sqrt{V}\bar{\lambda}^{\alpha}_a ({\gamma}^\mu)_{\alpha}^{\;\;\beta}\partial_\mu \lambda^a_{\beta}=2\I R\; \im \tau\int_{\mathbb{R}^3} \de^3x\; \sqrt{V}\bar{\lambda}_a {\gamma}^\mu \partial_\mu \lambda^a. 
\end{split}\label{GH3d-dual-omega}
\end{equation}
where $\mu,\nu=1,2,3$. As $r \to \infty$, the periodic scalars $\gamma$ and $\sigma$ obey the boundary condition
\begin{equation}
\gamma \to \frac{\theta_m}{4\pi R},\; \sigma \to \frac{\theta_e + 2n\pi}{4\pi R},\; n \in \Z.
\end{equation}

The rules for SUSY for the above action are as follows:
\begin{equation}
\begin{split}
&\delta \gamma=\frac{1}{\sqrt{V}} \tau\zeta_a \bar{\lambda}^a ,\\
& \delta \sigma=\frac{1}{\sqrt{V}}\zeta_a \bar{\lambda}^a,\\
 &\delta A_u =-\zeta_{a}\lambda^{a}+\frac{\epsilon R}{\sqrt{V}}\zeta_a \bar{\lambda}^a,\\
 &\delta A_v =-\zeta_{a}\lambda^{a},\\
&\delta \lambda_{\alpha}^{a} = -(\im \tau)^{-1} \partial_\mu (\gamma - \tau \;\sigma)(\gamma^{\mu})_{\alpha}^{\;\;\beta}\zeta_{\beta}^{ a},\\
&\delta \bar{\lambda}^a_{\alpha} =\frac{\I}{\sqrt{V}}(\gamma^{\mu})_{\alpha}^{\;\;\beta} {\zeta}^{a}_{\beta}\partial_\mu( A_{u}- A_{v}- \epsilon R \sigma).
\end{split} \label{susydual3D-3DspinorsGH-omega}
\end{equation}
This $\Omega$-deformed theory, reduced to 3d and dualized, can be readily identified as a generalized \hk sigma model, as we describe in the next subsection.

\subsection{Relation with the \hk sigma model}
Recall the general form of the bosonic part of the generalized \hk sigma model action.
\begin{equation}
S_b= \frac{1}{8 \pi} \int_{\mathbb{R}^3} \de^3x \left[ g_{ij} \partial_{\mu} \varphi^i \partial^{\mu} \varphi^j +  \epsilon_{\mu\nu\rho} G^{\mu\nu} \partial^{\rho}\varphi^i \cA_i\right]
\end{equation}
From equation \eqref{GH3d-dual-omega}, one can easily read off the metric $g$ and the $1$-form $\cA$.

\begin{equation}
\begin{split}
&g_{\gamma \gamma}=\frac{8 \pi R V}{ \im \tau}, \; g_{\sigma \sigma}=8 \pi R(\frac{ V \tau \bar{\tau}}{ \im \tau}+ \epsilon^2 R^2 \im \tau), \; g_{\gamma \sigma}=-\frac{4 \pi R V (\tau +\bar{\tau})}{ \im \tau},\\
&g_{uu}=8 \pi R \im \tau,\; g_{\sigma u}= -(\epsilon R) 8 \pi R \im \tau,\\
&g_{vv}=-8 \pi R \im \tau.\\
&\cA_{\sigma}=-8 \pi \I R^2 \gamma, \cA_{\gamma}=0, \cA_{u}=0, \cA_{v}=0.\\
& \F_{\gamma \sigma}=- 8 \pi \I R^2.
\end{split}
\end{equation}
The components of the inverse metric are
\begin{equation}
\begin{split}
&g^{\gamma \gamma}=\frac{1}{8\pi R V} \frac{\tau \bar{\tau}}{\im \tau}, g^{\gamma \sigma}=\frac{\tau+\bar\tau}{16\pi R V \im \tau}, g^{\gamma u}=\frac{\epsilon R(\tau+\bar\tau)}{16\pi R V \im \tau},\\
&g^{\sigma\sigma}=\frac{1}{8\pi R V \im \tau},g^{\sigma u}=\frac{\epsilon R}{8\pi R V \im \tau},\\
&g^{uu}=\frac{V + \epsilon^2 R^2}{8\pi R V \im \tau},g^{vv}=-\frac{1}{8\pi R \im \tau}.
\end{split}
\end{equation}

Define $N = -4 \pi R \im \tau, \tilde{N}=-\frac{4 \pi V}{R \im \tau}$. $Sp(r)$ indices (primed indices) are raised and lowered by the antisymmetric pairing
\begin{equation}
\begin{split}
\epsilon_{A' B'} = \begin{pmatrix} 0 & N \\ -N & 0 \end{pmatrix},\; \epsilon^{A' B'} = \begin{pmatrix} 0 & -\frac{1}{N} \\ \frac{1}{N} & 0 \end{pmatrix}
\end{split}
\end{equation}
while the unprimed indices are raised and lowered by the antisymmetric pairing
\begin{equation}
\begin{split}
\epsilon_{A B} = \begin{pmatrix} 0 & -1 \\ 1 & 0 \end{pmatrix},\; \epsilon^{A' B'} = \begin{pmatrix} 0 & 1 \\ -1 & 0 \end{pmatrix}
\end{split}
\end{equation}

The intertwiner $e$ can be explicitly written as
\begin{equation}
\begin{split}
&e^{AA'}_i \de \varphi^i=\begin{pmatrix} \de A_u -\epsilon R \de \sigma - \de A_v & \I \tilde{N} \de \gamma -\I \bar{\tau}\tilde{N} \de \sigma  \\ -\I \tilde{N} \de \gamma +\I {\tau}\tilde{N} \de \sigma & -\de A_u +\epsilon R \de \sigma -\de A_v \end{pmatrix}, \\
&e_{i\;AA'}\de \varphi^i=N\begin{pmatrix} \de A_u -\epsilon R \de \sigma+ \de A_v  & -\I \tilde{N} \de \gamma +\I {\tau}\tilde{N} \de \sigma  \\  \I \tilde{N} \de \gamma -\I \bar{\tau}\tilde{N} \de \sigma & -\de A_u +\epsilon R \de \sigma + \de A_v \end{pmatrix},\\
&e^{AA'}_0=0,\; e_{0AA'}=0.
\end{split} 
\end{equation}
The fermions and SUSY parameters can now be easily related:
\begin{equation}
\begin{split}
& \lambda^a_{\alpha} =\frac{1}{\sqrt{2V}} \left(\psi^{2'}_\alpha,\bar \psi_{\alpha}^ {2'} \right),\\
& \bar{\lambda}^a_{\alpha}=\frac{1}{\sqrt{2}} \left(\psi^{1'}_\alpha, \bar \psi_{\alpha}^ {1'} \right),\\
& \zeta^a_{\alpha}= \frac{\sqrt{V}}{\sqrt{2}} \left(-\zeta^2_{\alpha}, -\bar{\zeta}^2_{\alpha}\right),\\
&\zeta^1_{\sigma}=0, \; \bar{\zeta}^1_{\sigma}=0.
\end{split}
\end{equation}

In the sigma model, the constraint on the supersymmetry parameters $(\zeta^\alpha_A, \bar\zeta_\alpha^A)$ remain the same as before, namely
\begin{equation}
\begin{split}
&c^A \zeta^\alpha_A=0,\; c_A \bar\zeta_\alpha^A=0;\\
&c^A=(0,1), \;c_A=(1,0).
\end{split}
\end{equation}

Since  $\zeta^a_{\alpha}$ is a constant spinor, the above identification immediately implies
\begin{equation}
\begin{split}
&\partial_{\mu} \zeta^E + \frac{1}{2} \frac{\partial_{\mu} V}{V}\zeta^E=0,\\
&\partial_{\mu} \bar{\zeta}^E + \frac{1}{2} \frac{\partial_{\mu} V}{V}\bar{\zeta}^E=0,\\
\implies & f(\varphi^0)=\frac{R}{2V}.
\end{split}
\end{equation}

The effective $\tilde{q}^{A'}_{0B'}$ that appears in the extended \hk identities
will be given as
\begin{equation}
\begin{split}
\tilde{q}^{A'}_{0 B' }=\frac{R}{V}\begin{pmatrix}  1 & 0  \\ 0 & 0\end{pmatrix}
\end{split}
\end{equation}
The components of the connection $\Gamma^i_{jk}$ can now be derived using the extended \hk identities and can be shown to be the same as the undeformed case. One can also check that the equation for supersymmetry for a generalized \hk sigma model is satisfied for the metric, intertwiner and connection $A$ corresponding to the $\Omega$-deformed theory, as expected.
      
       Recall that in the undeformed theory curvature $\F$ of the connection $\cA$ was a $(1,1)$ form with respect to the complex structure specified by $c^A$. One can readily check that $\F$ is a $(1,1)$ form in the deformed theory as well by showing the $(2,0)$ component of $\F$ vanishes. Note that
\begin{equation}
\begin{split}
F^{(2,0)} =F^{ij}e_{i\; AA'} e_{j\; BB'}c^A c^B = F^{ij} e_{i\; 2A'} e_{j\; 2B'}
\end{split}
\end{equation}
which naturally vanishes for $A'=B'$ due to the antisymmetry of $F^{ij}$. For $A' \neq B'$, we have
\begin{equation}
\begin{split}
&A'=1',B'=2'\; : \;F^{ij} e_{i\; 21'} e_{j\; 22'} =F^{\gamma\sigma} e_{\gamma\;21'}e_{\sigma\; 22'}+F^{\gamma u} e_{\gamma 21'} e_{u 22'}=0, \\
&A'=2',B'=1'\; : \; F^{ij} e_{i\; 22'} e_{j\; 22'}=-F^{\gamma\sigma}e_{\sigma\; 22'} e_{\gamma\;21'} - F^{\gamma u} e_{u 22'}  e_{\gamma 21'} =0. 
\end{split}
\end{equation}

The \hk sigma model obtained by simply dimensionally reducing the $\Omega$-deformed theory on a Gibbons-Hawking space is asymptotically different from the sigma model in the undeformed case. To recast the $\Omega$-deformed theory such that it asymptotically looks equivalent to the undeformed one, we need to introduce the Nekrasov-Witten change of variables which we discuss in the next two sections.

\section{Nekrasov-Witten Change of Variables on $\mathbb{R}^3 \times S^1$}
\subsection{NW transformation: Usual 4d presentation}

Consider first the known example of  $U(1)$ SYM on $\mathbb{R}^3\times S^1$ $\Omega$-deformed using the Killing vector for the $U(1)$ isometry along the circle direction. The 4D action can be obtained from the action of $\mathcal{N}=1, U(1)$ SYM on the $\Omega$-deformed 6D background via dimensional reduction.
\begin{equation}
 \begin{split}
 S=& \frac{1}{g^2_{YM} L^2} \; \int d^6 x \sqrt{g_{\Omega}} \left[ \mathcal{L}_b +\mathcal{L}_f \right],\\
 \mathcal{L}_b=& \frac{1}{2} F_{mn} F_{mn} +F^2_{3u}+\sum_i\left( F_{iu}+\epsilon R F_{iy}\right)^2 - \sum_m F^2_{mv} -\left(F_{uv} + \epsilon R  F_{yv} \right)^2 ,\\
 \mathcal{L}_f =&\bar{\psi}_a \Gamma_m \partial_m \psi^a + \bar{\psi}_a \Gamma_4 \partial_u \psi^a +\bar{\psi}_a \Gamma_5 \partial_v \psi^a +\epsilon R \bar{\psi}_a \Gamma_4 \partial_y \psi^a.
 \end{split}
\end{equation}
where the fermionic field $\psi_a$ is a symplectic Majorana-Weyl spinor which transforms as a doublet of the
$SU(2)_R$ symmetry (the indices $m,n=0,1,2,3$ while $i,j=0,1,2$).

These SUSY rules for the $\Omega$-deformed action are
\begin{equation}
\begin{split}
&\delta A_i = -\bar{\psi}_a \Gamma_i  \zeta^a, \\
&\delta A_y = -\bar{\psi}_a \Gamma_3  \zeta^a, \\
&\delta A_u = -\bar{\psi}_a \Gamma_4  \zeta^a + \epsilon R \bar{\psi}_a \Gamma_3  \zeta^a,\\
& \delta A_v = -\bar{\psi}_a \Gamma_5  \zeta^a,\\
&\delta \psi^a= -\frac{1}{2} F_{mn} \Gamma_{mn} \zeta^a- F_{3u} \Gamma_{34} \zeta^a- (F_{iu}+ \epsilon R F_{iy})\Gamma_{i4} \zeta^a+F_{mv} \Gamma_{m5} \zeta^a+(F_{uv}+\epsilon R F_{yv} ) \Gamma_{45} \zeta^a,\\
&\delta \bar{\psi}^a= -\frac{1}{2} F_{mn}\bar{\zeta}^a \Gamma_{mn} - F_{3u} \bar{\zeta}^a\Gamma_{34} - (F_{iu}+ \epsilon R F_{iy})\bar{\zeta}^a\Gamma_{i4} +F_{mv} \bar{\zeta}^a\Gamma_{m5} +(F_{uv}+\epsilon R F_{yv} ) \bar{\zeta}^a\Gamma_{45} .\\
\end{split} 
\end{equation}

Now, consider combining the $\Omega$- deformation with a field redefinition which acts as $S: \;A_u \to \frac{A_u}{\sqrt{1+\epsilon^2 R^2}}$. $\Omega$-deformation combined with this field redefinition can be understood as a formal substitution in the undeformed action and the undeformed SUSY transformation, namely
\begin{equation}
\Omega \circ S :\;\; A_u \to \frac{A_u +\epsilon R A_y}{\sqrt{1+\epsilon^2 R^2}}.
\end{equation}
Generally, at every point on a given 4-manifold where the action of the $U(1)$ isometry is free, there exists a change of variables which can restore the deformed action (and SUSY transformations) to the undeformed one up to some rescaling of parameters. Since the $U(1)$ action does not have a fixed point on $\R^3 \times S^1$, $\Omega$-deformation may be ``negated" everywhere by the following change of variables:
\begin{equation}
NW:\;\;A_y \to \frac{A_y-\epsilon R A_u}{\sqrt{1+\epsilon^2 R^2}} \label{NWboson}.
\end{equation}
Note that $A_y$ is the gauge field in the circle direction. Therefore, while a shift of $A_y$ by a non-periodic scalar is a simple change of variables, one is usually not allowed to rescale $A_y$ in the manner shown above. However, as discussed in \cite{Nekrasov:2010ka}, this rescaling of $A_y$ can be understood as the rescaling of the $S^1$ radius $R$ and results in rescaling of the gauge coupling $g_{YM}$. As we will demonstrate momentarily, these rescalings appear naturally if one works with the dualized action in 3D.\\

Formally, the combination of the $\Omega$-deformation and the change of variables may be represented as a rotation in the $u - y$ plane:
\begin{equation}
\begin{split}
&A_u \to \cos{\theta} A_u + \sin{\theta} A_y,\\
&A_y \to \cos{\theta} A_y -\sin{\theta} A_u.
\end{split}
\end{equation}
where the angle of rotation $\theta$ is related to the parameter of $\Omega$-deformation by a simple equation:
\begin{equation}
\cos{\theta}=\frac{1}{\sqrt{1+\epsilon^2 R^2}}.
\end{equation}

One can readily check that the bosonic action is invariant under $NW \circ \Omega \circ S$. To restore the fermionic action as well as the SUSY rules to their original form, the fermions and SUSY parameters in the theory (written as 6D spinors) should transform in the following way \cite{Nekrasov:2010ka}:
\begin{equation}
\begin{split}
&\psi_a \to \exp{\left(\frac{\theta}{2} \Gamma_{34}\right)} \psi_a,\\
&\zeta_a \to \exp{\left(\frac{\theta}{2} \Gamma_{34}\right)} \zeta_a.
\end{split} 
\end{equation}

The $\Omega$-deformed 3D action, before dualization, is simply given by setting $V=1,B=0$ in \eqref{3Dactionb4dual} and may be obtained by dimensional reduction of the 4D action down to 3D. 
\begin{equation}
\begin{split}
&\mathcal{L}_b= \frac{1}{2}(F^{(3)}_{\mu \nu})^2 + (\partial_\mu \sigma)^2 +  \left(\partial_\mu A_u -\epsilon R\partial_\mu \sigma \right)^2 -  \left( \partial_\mu A_v \right)^2,\\
&\mathcal{L}_f= 2\I  \bar{\lambda}_a \gamma^\mu \partial_\mu \lambda^a.
\end{split}
\end{equation}
where $F^{(3)} =dA^{(3)}$. The SUSY rules for the deformed 3D theory are obtained via dimensional reduction of the 6D rules.
\begin{equation}
\begin{split}
&\delta A_\mu= -\I \left({\zeta}_a {\gamma}_{\mu}\bar{\lambda}^{a}-\bar{\zeta}_a {\gamma}_{\mu}\lambda^{a}\right)\; (\mu=1,2,3),\\
&\delta \sigma=(\zeta_a \bar{\lambda}^a+\bar{\zeta}_a \lambda^a),\\
& \delta A_u =-( \zeta_{a}\lambda^{a}-\bar{\zeta}_a\bar{\lambda}^a)+\epsilon R (\zeta_a \bar{\lambda}^a+\bar{\zeta}_a \lambda^a),\\
&\delta A_v =-(\zeta_{a}\lambda^{a}+\bar{\zeta}_a\bar{\lambda}^a), \\
&\delta \lambda^a_{\alpha} = -\frac{\I}{2} F_{\mu \nu}\epsilon^{\mu \nu \kappa}(\gamma_{\kappa})_{\alpha}^{\;\;\beta}\zeta_{\beta}^{a} +\I \partial_\mu\sigma (\gamma^\mu)_{\alpha}^{\;\;\beta}\zeta_{\beta}^{a}+\I (\gamma^{\mu})_{\alpha}^{\;\;\beta} \bar{\zeta}^a_{\beta}\partial_\mu (A_u+A_v-\epsilon R \sigma),  \\
& \delta \bar{\lambda}^a_{\alpha} =-\frac{\I}{2} F_{\mu \nu}\epsilon^{\mu \nu \kappa}(\gamma_{\kappa})_{\alpha}^{\;\;\beta}\bar{\zeta}_{\beta}^{ a} -\I \partial_\mu\sigma (\gamma^\mu)_{\alpha}^{\;\;\beta}\bar
{\zeta}_{\beta}^{a}+ \I(\gamma^{\mu})_{\alpha}^{\;\;\beta} {\zeta}^{a}_{\;\beta}\partial_\mu (A_u-A_v-\epsilon R \sigma).
\end{split} \label{susy3D-3Dspinorsflat}
\end{equation}

In 3D, the transformation $NW \circ \Omega \circ S $ amounts to transforming the bosonic fields $A_u, \sigma$ (but not $A_\mu$) as well as the fermionic fields $\lambda_a, \bar{\lambda}_a$ and the SUSY parameters $\zeta_a, \bar{\zeta}_a$ in the following fashion. 
\begin{equation}
\begin{split}
&A_u \to \frac{A_u -\epsilon R \sigma}{\sqrt{1+\epsilon^2 R^2}}=\cos{\theta} A_u -\sin{\theta} \sigma,\\
&\sigma \to \frac{\sigma+\epsilon R A_u}{\sqrt{1+\epsilon^2 R^2}}=\cos{\theta} \sigma +\sin{\theta} A_u,\\
&\lambda_a \to \cos{\frac{\theta}{2}}\lambda_a +\sin{\frac{\theta}{2}} \bar{\lambda}_a, \\
& \bar{\lambda}_a \to \cos{\frac{\theta}{2}}\bar{\lambda}_a -\sin{\frac{\theta}{2}} {\lambda}_a ,\\
&\zeta_a \to \cos{\frac{\theta}{2}}\zeta_a+\sin{\frac{\theta}{2}} \bar{\zeta}_a, \\
& \bar{\zeta}_a \to \cos{\frac{\theta}{2}}\bar{\zeta}_a -\sin{\frac{\theta}{2}} {\zeta}_a.\\
\end{split} 
\end{equation}
One can check that the bosonic action, the fermionic action as well as the supersymmetry transformations are invariant under the above transformation (up to total derivatives in the fermionic action).

Therefore, the action and the SUSY transformation rules written in terms of the redefined variables $A_u,\sigma, \lambda_a,\bar{\lambda}_a$ and the redefined supersymmetry parameters $\zeta_a,\bar{\zeta}_a$ are precisely the same as those for the undeformed theory. This is expected since action of the $U(1)$ isometry group is free at every point on $\R^3 \times S^1$ -- the Nekrasov-Witten change of variables can undo the $\Omega$-deformation at every point on this 4-manifold. 

\subsection{Dualization of the 3D action}
Let us explain the dualization of the 3D action obtained via the $NW\circ \Omega \circ S$ operation. So far we have not taken into account the usual topological $\theta$-term or a term involving the monopole number of the gauge field \cite{Dey:2014lja} in the analysis. Let us add these terms to the 4D action.
\begin{equation}
\Delta S^{(4)}= \I \frac{\re\tau}{4\pi}\int_{\mathbb{R}^3 \times S^1} F^{(4)}\wedge F^{(4)} + \I \frac{\theta_m}{4\pi^2 R} \int_{{S^2}_{\infty} \times S^1} R \de \chi \wedge F^{(4)}
\end{equation}

The first term is unaffected by the $\Omega$-deformation but transforms non-trivially under NW change of variables.
\begin{equation}
\begin{split}
\I  \int_{\mathbb{R}^3 \times S^1} F^{(4)} \wedge F^{(4)} \xrightarrow{dim. red.} & \I R \int_{\mathbb{R}^3} 2F^{(3)}\wedge \de \sigma \\
\xrightarrow{NW} & \frac{\I R\;\re\tau}{\sqrt{1+\epsilon^2R^2}} \int_{\mathbb{R}^3} 2F^{(3)}\wedge \de (\sigma+\epsilon R A_u)\\
=& \frac{\I R\;\re\tau}{\sqrt{1+\epsilon^2R^2}} \int_{\mathbb{R}^3} (2F^{(3)}\wedge \de \sigma + \epsilon R \de (A_u F^{(3)} ))
\end{split}
\end{equation}
Note that the adjoint scalar $A_u$ is globally defined while $A^{(3)}$ and $\sigma$ (being a periodic scalar) are not. Therefore one cannot write the first term as a total derivative. However, the second term clearly can be written as total derivative. This implies that we can  safely drop the second contribution from the topological term.

To dualize the action, it is convenient to redefine the gauge coupling in the following fashion
\begin{equation}
\begin{split}
&\im \hat{\tau}=\im \tau \sqrt{1+\epsilon^2 R^2} \implies \hat{g}^2_{YM} =\frac{g^2_{YM}}{\sqrt{1+\epsilon^2R^2}}\\
&\re \hat{\tau}= \re \tau
\end{split}
\end{equation}

Therefore, the 3D action may be rewritten as 
\begin{equation}
\begin{split}
S^b_{(3)}=&\frac{R \;\im{\hat \tau}}{\sqrt{1+\epsilon^2R^2}} \int_{\mathbb{R}^3} \de^3x \left[\frac{1}{2}F^{(3)}_{\mu\nu}F^{(3)\;\mu\nu} +\partial^\mu\sigma \partial_\mu \sigma+ \partial^\mu A_u \partial_\mu A_u- \partial^\mu A_v \partial_\mu A_v\right]\\
&+ \frac{\I R\; \re {\hat \tau}}{\sqrt{1+\epsilon^2R^2}}  \int_{\mathbb{R}^3} \de^3x \epsilon^{\mu\nu\kappa} F^{(3)}_{\mu\nu} \partial_\kappa \sigma+\I \frac{\theta_m}{2\pi}\int_{S^2_{\infty}}F^{(3)} \\
S^f_{(3)}=& \frac{2\I R\; \im \hat\tau}{\sqrt{1+\epsilon^2R^2}}\int_{\mathbb{R}^3} \de^3x\;  \bar{\lambda}^{\alpha}_a ({\gamma}^\mu)_{\alpha}^{\;\;\beta}\partial_\mu\lambda^a_{\beta}=\frac{2\I R\; \im \hat\tau}{\sqrt{1+\epsilon^2R^2}} \int_{\mathbb{R}^3} \de^3x\; \bar{\lambda}_a {\gamma}^\mu \partial_\mu\lambda^a
\end{split}
\end{equation}
One can now add the dualizing term $S_{dual}= -\frac{2\I R}{\sqrt{1+\epsilon^2R^2}}\int_{\mathbb{R}^3} \gamma \wedge \de F^{(3)}$ which imposes the Bianchi identity $dF^{(3)}=0$. The equation of motion for $F^{(3)}$ gives
\begin{equation}
F^{(3)} =-\I (\im \hat\tau)^{-1} \star \de (\gamma -(\re \hat\tau)\sigma)
\end{equation}
Integrating out $F^{(3)}$ as a Lagrange multiplier, one obtains the following dualized 3D action:
\begin{equation}
\begin{split}
&S^b_{(3)}=R' \int_{\mathbb{R}^3} \de^3x \left[(\im \hat\tau)^{-1}|\de\gamma-\hat\tau \de\sigma|^2  +(\im \hat\tau) (\partial^i A_u \partial_i A_u- \partial^i A_v \partial_i A_v)\right]\\
&S^f_{(3)}= 2\I R' \; \im \hat\tau \int_{\mathbb{R}^3} \de^3x\; \bar{\lambda}_a {\gamma}^i\partial_i\lambda^a .
\end{split} 
\end{equation}
where $R'=\frac{R}{\sqrt{1+\epsilon^2R^2}}$ is the rescaled radius. The boundary condition on the periodic scalars as $r \to \infty$ are
\begin{equation}
\gamma \to \frac{\theta_m}{4\pi R'},\; \sigma \to \frac{\theta_e + 2n\pi}{4\pi R'}
\end{equation}
where $n \in \Z$.

The SUSY transformation rules of the dualized 3D action are
\begin{equation}
\begin{split}
&\delta \gamma= ({\hat \tau} \zeta_a \bar{\lambda}^a+\bar{\hat \tau}\bar{\zeta}_a \lambda^a),\\
&\delta \sigma=(\zeta_a \bar{\lambda}^a+\bar{\zeta}_a \lambda^a),\\
& \delta A_u =-( \zeta_{a}\lambda^{a}-\bar{\zeta}_a\bar{\lambda}^a),\;\delta A_v =-(\zeta_{a}\lambda^{a}+\bar{\zeta}_a\bar{\lambda}^a), \;\\
&\delta \lambda_{\alpha}^{a} = -(\im \hat \tau)^{-1} (\partial_k\gamma-{\hat \tau} \partial_k\sigma)(\gamma^{k})_{\alpha}^{\;\;\beta}\zeta_{\beta}^{ a} + \I (\gamma^{i})_{\alpha}^{\;\;\beta} \bar{\zeta}^a_{\beta}(\partial_i A_{u}+ \partial_i A_{v}),\\
& \delta \bar{\lambda}^a_{\alpha} =-(\im \hat \tau)^{-1} (\partial_k\gamma-\bar{\hat \tau} \partial_k\sigma)(\gamma^{k})_{\alpha}^{\;\;\beta}\bar{\zeta}_{\beta}^{ a}+ \I (\gamma^{i})_{\alpha}^{\;\;\beta} {\zeta}^{a}_{\beta}(\partial_i A_{u}- \partial_i A_{v}).
\end{split} 
\end{equation}

 The 3D theory derived above is therefore exactly the same as the 3D theory obtained by dimensional reduction of the undeformed theory on $\mathbb{R}^3 \times S^1$, with $\tau$ being replaced by $\hat\tau$ and $R$ replaced by $R'$. As expected in $\R^3 \times S^1$, the $NW \circ \Omega \circ S$-transformed theory is equivalent to the undeformed theory up to rescalings of the radius $R$ and the gauge coupling $g_{YM}$.

\subsection{Alternative description of NW from the dualized 3D action} \label{alt-NW}
Now, we present an alternative way to describe the NW transformation. In this approach, one can understand the transformation $NW \circ \Omega \circ S$ purely in terms of the dualized 3D action -- the generalized \hk sigma model-- without having to resort to the 4D/6D picture. We demonstrate that one obtains the same result as above. The crucial step in this approach is to derive how the scalar $\gamma$ should transform under $NW\circ \Omega \circ S$.\\

Recall the ``undeformed" dualized 3D action: 
\begin{equation}
\begin{split}
{S}^b_{(3)} &= R \int_{\mathbb{R}^3}\Big[ (\im \tau)^{-1} |\partial_\mu\gamma - \tau \partial_\mu \sigma|^2+ \im \tau (\partial_\mu A_u)^2 
-  \im \tau (\partial_\mu A_v)^2\Big]\\
{S}^f_{(3)} &= 2\I R\; \im \tau\int_{\mathbb{R}^3} \de^3x\;  \bar{\lambda}^{\alpha}_a ({\gamma}^\mu)_{\alpha}^{\;\;\beta}\partial_\mu\lambda^a_{\beta}=2\I R\; \im \tau\int_{\mathbb{R}^3} \de^3x\; \bar{\lambda}_a {\gamma}^\mu\partial_\mu\lambda^a. 
\end{split}\label{flat3d-dual}
\end{equation}
with the following rules for SUSY:
\begin{equation}
\begin{split}
&\delta \gamma= (\tau \zeta_a \bar{\lambda}^a+\bar{\tau}\bar{\zeta}_a \lambda^a),\\
&\delta \sigma=(\zeta_a \bar{\lambda}^a+\bar{\zeta}_a \lambda^a),\\
& \delta A_u =-( \zeta_{a}\lambda^{a}-\bar{\zeta}_a\bar{\lambda}^a),\;\delta A_v =-(\zeta_{a}\lambda^{a}+\bar{\zeta}_a\bar{\lambda}^a), \;\\
&\delta \lambda_{\alpha}^{a} = -(\im \tau)^{-1} (\partial_\mu \gamma-\tau \partial_\mu \sigma)(\gamma^{\mu})_{\alpha}^{\;\;\beta}\zeta_{\beta}^{a} + \I (\gamma^{\mu})_{\alpha}^{\;\;\beta} \bar{\zeta}^a_{\beta}(\partial_\mu A_{u}+ \partial_\mu A_{v}),\\
& \delta \bar{\lambda}^a_{\alpha} =-(\im \tau)^{-1} (\partial_\mu\gamma-\bar{\tau} \partial_\mu\sigma)(\gamma^{\mu})_{\alpha}^{\;\;\beta}\bar{\zeta}_{\beta}^{ a}+\I (\gamma^{\mu})_{\alpha}^{\beta} {\zeta}^{a}_{\;\beta}(\partial_\mu A_{u}- \partial_\mu A_{v}).
\end{split} \label{susy3D-3Dspinors-dual}
\end{equation}

The transformation $NW \circ \Omega \circ S$ acts on the fields $\sigma, A_u,  \lambda_{\alpha}^{a}, \bar{\lambda}^a_{\alpha}$ and the SUSY parameters as follows:
\begin{equation}
\begin{split}
&A_u \to \frac{A_u -\epsilon R \sigma}{\sqrt{1+\epsilon^2 R^2}}=\cos{\theta} A_u -\sin{\theta} \sigma,\\
&\sigma \to \frac{\sigma+\epsilon R A_u}{\sqrt{1+\epsilon^2 R^2}}=\cos{\theta} \sigma +\sin{\theta} A_u,\\
&\lambda_a \to \cos{\frac{\theta}{2}}\lambda_a +\sin{\frac{\theta}{2}} \bar{\lambda}_a, \\
& \bar{\lambda}_a \to \cos{\frac{\theta}{2}}\bar{\lambda}_a -\sin{\frac{\theta}{2}} {\lambda}_a ,\\
&\zeta_a \to \cos{\frac{\theta}{2}}\zeta_a+\sin{\frac{\theta}{2}} \bar{\zeta}_a, \\
& \bar{\zeta}_a \to \cos{\frac{\theta}{2}}\bar{\zeta}_a -\sin{\frac{\theta}{2}} {\zeta}_a.\\
\end{split} \label{NWOSfields}
\end{equation}
Recall that the dual photon $\gamma$ is related to the curvature $F^{(3)}$ via the dualization condition: 
\begin{equation}
F^{(3)} =-\I (\im \tau)^{-1} \star \de (\gamma -(\re \tau)\sigma)
\end{equation}

Since $F^{(3)}$ is invariant under NW, demanding that $\gamma$ transforms as 
\begin{empheq}[box=\fbox]{gather}
\gamma \to \cos{\theta} \gamma + (\re \tau \sin{\theta}) A_u \label{NWOSgamma}
\end{empheq}
and defining a rescaled coupling constant $\hat\tau$ as
\begin{equation}
\begin{split}
&\im \tau=\im \hat{\tau}\cos{\theta}\\
&\hat{\tau}=\re \tau + \I \im \hat{\tau}.
\end{split} 
\end{equation}
the dualization condition in terms of the ``new'' fields $\gamma$ and $\sigma$ assumes the form
\begin{equation}
F^{(3)} =-\I (\im \hat \tau)^{-1} \star \de (\gamma -(\re \hat\tau)\sigma) \label{newdualization}
\end{equation}

Applying the transformations \eqref{NWOSfields} and \eqref{NWOSgamma} to the undeformed theory defined by \eqref{flat3d-dual} and \eqref{susy3D-3Dspinors-dual} and the using the dualization condition \eqref{newdualization}, we obtain 
\begin{equation}
\begin{split}
&S^b_{(3)}=R' \int_{\mathbb{R}^3} \de^3x \left[(\im \hat{\tau})^{-1} |\partial_\mu\gamma - \hat{\tau} \partial_\mu \sigma|^2+ \im \hat{\tau} (\partial_\mu A_u)^2 -  \im \hat{\tau} (\partial_\mu A_v)^2\right]\\
&S^f_{(3)}= 2\I R'\; \im \hat\tau \int_{\mathbb{R}^3} \de^3x\; \bar{\lambda}_a {\gamma}^\mu\partial_\mu\lambda^a, \;\; R'=R\cos{\theta}.
\end{split} 
\end{equation}
with the following rules of SUSY
\begin{equation}
\begin{split}
&\delta \gamma= ({\hat \tau} \zeta_a \bar{\lambda}^a+\bar{\hat \tau}\bar{\zeta}_a \lambda^a),\\
&\delta \sigma=(\zeta_a \bar{\lambda}^a+\bar{\zeta}_a \lambda^a),\\
& \delta A_u =-( \zeta_{a}\lambda^{a}-\bar{\zeta}_a\bar{\lambda}^a),\\
&\delta A_v =-(\zeta_{a}\lambda^{a}+\bar{\zeta}_a\bar{\lambda}^a), \\
&\delta \lambda_{\alpha}^{a} = -(\im \hat \tau)^{-1} (\partial_\mu\gamma-{\hat \tau} \partial_\mu\sigma)(\gamma^{\mu})_{\alpha}^{\;\;\beta}\zeta_{\beta}^{ a} + \I (\gamma^{\mu})_{\alpha}^{\;\;\beta} \bar{\zeta}^a_{\beta}(\partial_\mu A_{u}+ \partial_\mu A_{v}),\\
& \delta \bar{\lambda}^a_{\alpha} =-(\im \hat \tau)^{-1} (\partial_\mu\gamma-\bar{\hat \tau} \partial_\mu\sigma)(\gamma^{\mu})_{\alpha}^{\;\;\beta}\bar{\zeta}_{\beta}^{ a}+ \I (\gamma^{\mu})_{\alpha}^{\;\;\beta} {\zeta}^{a}_{\beta}(\partial_\mu A_{u}- \partial_\mu A_{v}).
\end{split} 
\end{equation}

As expected, the dualized action in 3D is ``almost" invariant theory under the $NW\circ \Omega \circ S$ operation -- the radius R and the gauge coupling undergo a rescaling -- precisely as worked out in the original presentation of Nekrasov and Witten. In the next section, we apply this alternative approach to NW change of variables to $\Omega$-deformed theory on NUT space.

\section{Nekrasov-Witten Change of Variables on Gibbons-Hawking Space}
In this section we study in detail the 3D sigma model action obtained by dimensionally reducing an  $\Omega$-deformed and $NW$-transformed $\N=2, U(1)$ SYM theory on a Gibbons-Hawking space. The $U(1)$ isometry used to $\Omega$-deform the theory has a fixed point at the NUT center. Therefore the $NW$-transformed theory is equivalent (up to some rescaling of parameters) to the undeformed sigma model only in the asymptotic limit and differs from it as one approaches the NUT center. Nevertheless, the $NW \circ \Omega \circ S$-transformed theory can be understood as an example of a generalized \hk sigma model introduced in \cite{Dey:2014lja}, as we demonstrate below.

\subsection{NW from 3d dualized action}
We describe the $NW\circ \Omega \circ S$ transformation using the alternative approach outlined in the previous section. The starting point is the dualized 3D action of the undeformed theory -- i.e. the generalized \hk sigma model associated with $\N=2, U(1)$ SYM theory on a Gibbons-Hawking space.
\begin{equation}
\begin{split}
{S}^b_{(3)} &= R\int_{\mathbb{R}^3} \de^3 x \Big[V (\im \tau)^{-1} |\partial_\mu\gamma - \tau \partial_\mu \sigma|^2+ \im \tau (\partial_\mu A_u)^2 
-  \im \tau (\partial_\mu A_v)^2\Big]\\
&\;\;\;  -2\I R^2\int_{\mathbb{R}^3} \gamma \wedge \de\sigma \wedge dB,\\
{S}^f_{(3)} &= 2\I R\; \im \tau\int_{\mathbb{R}^3} \de^3x\; \sqrt{V}\bar{\lambda}_a {\gamma}^\mu\partial_\mu\lambda^a. 
\end{split}\label{GH3d-dual}
\end{equation}
with the following rules for SUSY:
\begin{equation}
\begin{split}
&\delta \gamma=\frac{1}{\sqrt{V}} \tau\zeta_a \bar{\lambda}^a,\\
& \delta \sigma=\frac{1}{\sqrt{V}}\zeta_a \bar{\lambda}^a,\\
&\delta A_u =-\zeta_{a}\lambda^{a},\\
&\delta A_v =-\zeta_{a}\lambda^{a}, \\
&\delta \lambda_{\alpha}^{a} = -(\im \tau)^{-1} \partial_\mu(\gamma - \tau \;\sigma)(\gamma^{\mu})_{\alpha}^{\;\;\beta}\zeta_{\beta}^{ a},\\
&\delta \bar{\lambda}^a_{\alpha} =\frac{\I}{\sqrt{V}}(\gamma^{\mu})_{\alpha}^{\;\;\beta} {\zeta}^{a}_{\beta}\partial_\mu( A_{u}- A_{v}).
\end{split} \label{susydual3D-3DspinorsGH}
\end{equation}
The above action is asymptotically equivalent to the dualized 3D theory obtained by dimensional reduction of a $4d, \N=2$ theory on $\R^3 \times S^1$.  Supersymmetry for this theory  is generated by an 8-complex dimensional space of SUSY parameters $(\zeta^a_\alpha,\bar{\zeta}^a_\alpha)$ with $a=1,2$ labeling the $SU(2)_R$ index.  Asymptotically, on a  Gibbons-Hawking space, the supersymmetry rules are given by those for the dimensionally reduced theory on $\R^3 \times S^1$ subject to the constraint
\begin{equation}
\bar\zeta^a_\alpha=0.
\end{equation}
thereby reducing the number of supersymmetries from 8 to 4.\\

Next, we study the action of $NW\circ\Omega\circ S$ transformation on this theory -- the action on the fields are the same as discussed in section \S\ref{alt-NW} for $\R^3 \times S^1$, namely
\begin{equation}
\begin{split}
&A_u \to \cos{\theta} A_u -\sin{\theta} \sigma,\\
&\sigma \to \cos{\theta} \sigma +\sin{\theta} A_u,\\
&\gamma \to \cos{\theta} \gamma + (\re \tau \sin{\theta}) A_u,\\
&\lambda_a \to \cos{\frac{\theta}{2}}\lambda_a +\sin{\frac{\theta}{2}} \bar{\lambda}_a, \\
&\bar{\lambda}_a \to \cos{\frac{\theta}{2}}\bar{\lambda}_a -\sin{\frac{\theta}{2}} {\lambda}_a.
\end{split} \label{NWOSfields-GH}
\end{equation}

The bosonic action under the above transformation becomes 
\begin{equation}
\begin{split}
&S_b=S_0 +S_{GH}\\
&S_0=R'\int_{\mathbb{R}^3} \de^3 x \Big[V (\im \hat\tau)^{-1} |\partial_\mu\gamma - \hat\tau \partial_\mu \sigma|^2+ \im \hat\tau ((\partial_\mu A_u)^2 
-  (\partial_\mu A_v)^2)\Big]\\
&+R'\im \hat\tau \int_{\mathbb{R}^3} \de^3 x (V-1) \Big(\sin^2{\theta} (\partial_\mu A_u)^2 - \sin^2{\theta} (\partial_\mu \sigma)^2+ \sin{2\theta} \partial_\mu A_u \partial_\mu \sigma \Big)\\
&S_{GH}=-\I R'^2 \int_{\mathbb{R}^3} \de^3 x \; \epsilon_{\mu \nu \rho} G_{\mu \nu} \Big(\gamma \partial_\rho \sigma + \re \tau \tan{\theta} A_u \partial_\rho \sigma  +\tan{\theta} \gamma \partial_\rho A_u + \re \tau \tan^2{\theta} A_u \partial_\rho A_u\Big) \\
\end{split} \label{NWOS-GH-bosaction}
\end{equation}
The fermionic action is
\begin{equation}
\begin{split}
S_f= &2\I R' \im \hat{\tau} \int_{\mathbb{R}^3} \de^3x\; \sqrt{V}\Big(\bar{\lambda}_a {\gamma}^\mu\partial_\mu\lambda^a + \frac{1}{4} \frac{\partial_\mu V}{V} (1-\cos{\theta})\bar{\lambda}_a {\gamma}^\mu\lambda^a \\
&-\frac{1}{4} \frac{\partial_\mu V}{V} \sin{\theta}\bar{\lambda}_a {\gamma}^\mu\bar{\lambda}^a+\frac{1}{4} \frac{\partial_\mu V}{V} \sin{\theta}{\lambda}_a {\gamma}^\mu{\lambda}^a\Big)
\end{split} \label{NWOS-GH-fermaction}
\end{equation}
Note that the above action is asymptotically (in the $r \to \infty$ limit) equivalent to the $NW \circ \Omega \circ S$-transformed theory on $\R^3 \times S^1$ but differs from the latter at finite $r$. Also, in the limit where the $\Omega$-deformation parameter vanishes (i.e. $\theta \to 0$), the action is restored to the generalized \hk sigma model given in \eqref{GH3d-dual}.\\

It is convenient to describe the SUSY of this $NW\circ\Omega\circ S$-transformed theory in terms of the parameters $\hat{\zeta}_a$ and $\hat{\bar{\zeta}}_a$ which are defined as
\begin{equation}
\hat{\zeta}_a= {\zeta}_a \cos{\frac{\theta}{2}}, \; \hat{\bar{\zeta}}_a= {{\zeta}}_a \sin{\frac{\theta}{2}}
\end{equation}
Note that we still have 4 supersymmetries but the SUSY parameters $(\hat{\zeta}_a,\hat{\bar{\zeta}}_a)$ obey a different constraint, namely
\begin{equation}
\hat{\bar{\zeta}}_a=\tan{\frac{\theta}{2}}\hat{\zeta}_a
\end{equation}

With this redefinition, the deformed SUSY transformation rules for scalars in the  $NW\circ\Omega\circ S$-transformed theory  can then be summarized as
\begin{equation}
\begin{split}
&\delta \gamma= ({\hat \tau} \hat{\zeta}_a \bar{\lambda}^a+\bar{\hat \tau}\hat{\bar{\zeta}}_a \lambda^a) +(\frac{1}{\sqrt{V}}-1) (\re \tau \cos{\theta}+ \I \im \hat\tau) ( \hat{\zeta}_{a}\bar{\lambda}^{a}-\hat{\bar{\zeta}}_a {\lambda}^a),\\
&\delta \sigma=\hat{\zeta}_a \bar{\lambda}^a+\hat{\bar{\zeta}}_a \lambda^a + (\frac{1}{\sqrt{V}}-1)\cos{\theta}( \hat{\zeta}_{a}\bar{\lambda}^{a}-\hat{\bar{\zeta}}_a {\lambda}^a) ,\\
& \delta A_u =-( \hat{\zeta}_{a}\lambda^{a}-\hat{\bar{\zeta}}_a\bar{\lambda}^a) +(\frac{1}{\sqrt{V}}-1)\sin{\theta}( \hat{\zeta}_{a}\bar{\lambda}^{a}-\hat{\bar{\zeta}}_a {\lambda}^a) ,\\
&\delta A_v =-(\hat{\zeta_{a}}\lambda^{a}+\hat{\bar{\zeta}}_a\bar{\lambda}^a).
\end{split} \label{def-susy-bos}
\end{equation}
while those for the fermions are 
\begin{equation}
\begin{split}
&\delta \lambda_{\alpha}^{a} = -(\im \hat \tau)^{-1} (\partial_\mu\gamma-{\hat \tau} \partial_\mu\sigma)(\gamma^{\mu})_{\alpha}^{\;\;\beta} \hat{\zeta}_{\beta}^{ a} + \I (\gamma^{\mu})_{\alpha}^{\;\;\beta} \hat{\bar{\zeta}}^a_{\beta}(\partial_\mu A_{u}+ \partial_\mu A_{v})\\
&+\I (\frac{1}{\sqrt{V}}-1) (1-\cos{\theta}) \partial_\mu \sigma (\gamma^{\mu})_{\alpha}^{\;\;\beta} \hat{\zeta}_{\beta}^{ a}-\I (\frac{1}{\sqrt{V}}-1) \partial_\mu(A_u \cos{\theta} -A_v)(\gamma^{\mu})_{\alpha}^{\;\;\beta} \hat{\bar{\zeta}}^a_{\beta},\\
& \delta \bar{\lambda}^a_{\alpha} =-(\im \hat \tau)^{-1} (\partial_\mu\gamma-\bar{\hat \tau} \partial_\mu \sigma)(\gamma^{\mu})_{\alpha}^{\;\;\beta}\hat{\bar{\zeta}}_{\beta}^{ a}+ \I (\gamma^{\mu})_{\alpha}^{\;\;\beta} \hat{\zeta}^{a}_{\beta}(\partial_\mu A_{u}- \partial_\mu A_{v})\\
&-\I (\frac{1}{\sqrt{V}}-1) (1+\cos{\theta}) \partial_\mu \sigma (\gamma^{\mu})_{\alpha}^{\;\;\beta} \hat{\bar{\zeta}}_{\beta}^{ a} + \I (\frac{1}{\sqrt{V}}-1)\partial_\mu(A_u \cos{\theta} -A_v)(\gamma^{\mu})_{\alpha}^{\;\;\beta} \hat{\zeta}^a_{\beta}.
\end{split} \label{def-susy-ferm}
\end{equation}
Again, in the asymptotic limit, the SUSY transformation has the same form as the $NW \circ \Omega \circ S$-transformed 3D theory obtained via dimensional reduction from $\R^3 \times S^1$ with a certain constraint on the 8-complex dimensional space of supersymmetry parameters $(\zeta^a_\alpha,\bar{\zeta}^a_\alpha)$. While the constraint for the undeformed case was simply $\bar{\zeta}^a_\alpha=0$, the corresponding constraint for the deformed theory is 
\begin{equation}
\sin{\frac{\theta}{2}}\hat{\zeta_{a}} = \cos{\frac{\theta}{2}} \hat{\bar{\zeta}}_a.
\end{equation}
This is basically related to the fact that although the asymptotic actions of the two theories have the same form, the preferred complex structures characteristic of the respective generalized \hk sigma models are different. 

In the limit $\theta \to 0$, it is clear that the above rules are restored to the undeformed version given in \eqref{susydual3D-3DspinorsGH} (recall that $\hat{\zeta}^{a}_{\alpha} \to 0$ in this limit), as one would expect.\\

Given \eqref{NWOS-GH-bosaction}- \eqref{def-susy-ferm}, one can now read off the various data of the generalized \hk sigma model -- which we  describe in the next subsection.

\subsection{Deformed theory as a \hk sigma model}
\subsubsection{Undeformed Theory}
To orient the reader with the basic computation, we briefly describe how the undeformed theory given by \eqref{GH3d-dual} -\eqref{susydual3D-3DspinorsGH} can be recast as a generalized \hk sigma model as reviewed in \S\ref{genhk-rev}. Recall that for a $U(1)$ SYM on Gibbons-Hawking background the target space of the sigma model $\ctM$ (which can be thought of as a 1-parameter family of \hk manifolds) restricts to the \hk manifold $\ctM[\varphi^0] \cong \C \times T^2$ parametrized by the scalars $\phi,\bar{\phi}$ and the periodic scalars $\sigma,\gamma$. In our notation, $\varphi^0$ is a harmonic scalar related to the function $V(\vec{x})$, i.e. $\varphi^0=\frac{V(\vec{x})}{R}$. 

Aside from the standard kinetic term for the scalars, the bosonic action also consists of a term involving the pull-back of a holomorphic connection $\cA$ on $\ctM$. The bosonic action derived via dimensional reduction \eqref{GH3d-dual}, therefore, allows one to read off the metric $g$ on $\ctM$ and the 1-form $\cA$.
 \begin{equation}
\begin{split}
&S_b= \frac{1}{8 \pi} \int_{\mathbb{R}^3} \de^3x \left[ g_{ij} \partial_{\mu} \varphi^i \partial^{\mu} \varphi^j +  \epsilon_{\mu\nu\rho} G^{\mu\nu} \partial^{\rho}\varphi^i \cA_i\right]\\
&g_{\gamma \gamma}=\frac{8 \pi R V}{ \im \tau}, \; g_{\sigma \sigma}=\frac{8 \pi R V \tau \bar{\tau}}{ \im \tau}, \; g_{\gamma \sigma}=-\frac{4 \pi R V (\tau +\bar{\tau})}{ \im \tau},\\
&g_{uu}=8 \pi R \im \tau,\; g_{vv}=-8 \pi R \im \tau.\\
&\cA_{\sigma}=-8 \pi \I R^2 \gamma, \cA_{\gamma}=0, \cA_{u}=0, \cA_{v}=0.\\
& \F_{\gamma \sigma}=- 8 \pi \I R^2.
\end{split}
\end{equation}
Let us define the following local coordinates on $\ctM$ : $y= \gamma - \tau \sigma, \bar{y}= \gamma - \bar{\tau} \sigma$. In the basis of local coordinates $\{y,\bar{y}, \phi, \bar{\phi}\}$, the holomorphic connection $\cA$ and its curvature $\F$  are given as
\begin{equation}
\begin{split}
&\mathcal{A}_y=-\mathcal{A}_{\bar{y}} =8 \pi \I \frac{(\tau \bar{y} - \bar{\tau} y)}{(\tau -\bar{\tau})^2}, \label{conn-1}\\
&\F_{y\bar{y}}=-\frac{8 \pi \I}{(\tau -\bar{\tau})}.
\end{split}
\end{equation}

The complexified tangent space of the \hk fiber admits the following decomposition $T_\C \ctM[\varphi^0] = H \otimes E$ where $H$ is 2-complex dimensional trivial vector bundle on $\ctM[\varphi^0]$ with a structure group $Sp(1)$ while $E$ is a 2-complex dimensional vector bundle with an $Sp(1)$ connection.\\

The fermions of the sigma model $\psi^{E'}_{\alpha}\;(E'=1,2)$ transform in the fundamental representation of $Sp(1)_E$ indicated by the primed indices which are raised and lowered by the antisymmetric pairing
\begin{equation}
\begin{split}
&\epsilon_{A' B'} = \begin{pmatrix} 0 & N \\ -N & 0 \end{pmatrix},\; \epsilon^{A' B'} = \begin{pmatrix} 0 & -\frac{1}{N} \\ \frac{1}{N} & 0 \end{pmatrix}\\
& N = -4 \pi R \im \tau, \; \tilde{N}=-\frac{4 \pi V}{R \im \tau}.
\end{split}
\end{equation}
 The supersymmetry parameters  $\zeta^E_{\alpha}, \bar{\zeta}^E_{\alpha}$ transform in the fundamental of $Sp(1)_H$ indicated by unprimed indices which are raised and lowered by the antisymmetric pairing
\begin{equation}
\begin{split}
\epsilon_{A B} = \begin{pmatrix} 0 & -1 \\ 1 & 0 \end{pmatrix},\; \epsilon^{A' B'} = \begin{pmatrix} 0 & 1 \\ -1 & 0 \end{pmatrix}
\end{split}
\end{equation}

From the supersymmetry rules in \eqref{susydual3D-3DspinorsGH}, the surjective map $e: T_\C \ctM \to H \otimes E$ can be explicitly written in local coordinates.
\begin{equation}
\begin{split}
&e^{AA'}_i \de \varphi^i=\begin{pmatrix} \de A_u - \de A_v & \I \tilde{N} \de \gamma -\I \bar{\tau}\tilde{N} \de \sigma  \\ -\I \tilde{N} \de \gamma +\I {\tau}\tilde{N} \de \sigma & -\de A_u -\de A_v \end{pmatrix}, \\
&e_{i\;AA'}\de \varphi^i=N\begin{pmatrix} \de A_u + \de A_v  & -\I \tilde{N} \de \gamma +\I {\tau}\tilde{N} \de \sigma  \\  \I \tilde{N} \de \gamma -\I \bar{\tau}\tilde{N} \de \sigma & -\de A_u + \de A_v \end{pmatrix},\\
&e^{AA'}_0=0,\; e_{0AA'}=0.
\end{split} 
\end{equation}
The fermions and SUSY parameters in the gauge theory can now be easily related to those in the sigma model. Recall that $a=1,2$ denotes the manifest $SU(2)_R$ index in the gauge theory.
\begin{equation}
\begin{split}
& \lambda^a_{\alpha} =\frac{1}{\sqrt{2V}} \left(\psi^{2'}_\alpha,\bar \psi_{\alpha}^ {2'} \right),\\
& \bar{\lambda}^a_{\alpha}=\frac{1}{\sqrt{2}} \left(\psi^{1'}_\alpha, \bar \psi_{\alpha}^ {1'} \right),\\
& \zeta^a_{\alpha}= \frac{\sqrt{V}}{\sqrt{2}} \left(-\zeta^2_{\alpha}, -\bar{\zeta}^2_{\alpha}\right),\\
&\zeta^1_{\sigma}=0, \; \bar{\zeta}^1_{\sigma}=0.
\end{split}
\end{equation}
In the generalized sigma model, the constraint on the supersymmetry parameters $(\zeta^\alpha_A, \bar\zeta_\alpha^A)$ may be written as 
\begin{equation}
\begin{split}
&c^A \zeta^\alpha_A=0,\; c_A \bar\zeta_\alpha^A=0\\
&c^A=(0,1), \;c_A=(1,0).
\end{split}
\end{equation}

Since  $\zeta^a_{\alpha}$ is a constant spinor, the above identification implies that $\zeta^E_{\alpha}, \bar{\zeta}^E_{\alpha}$ obey an equation of the form $\partial_{\mu} \zeta^E + f(\varphi^0) \partial_{\mu} \varphi^0 \zeta^E=0$, i.e.
\begin{equation}
\begin{split}
&\partial_{\mu} \zeta^E + \frac{1}{2} \frac{\partial_{\mu} V}{V}\zeta^E=0,\\
&\partial_{\mu} \bar{\zeta}^E + \frac{1}{2} \frac{\partial_{\mu} V}{V}\bar{\zeta}^E=0,\\
\implies & f(\varphi^0)=\frac{R}{2V}.
\end{split}
\end{equation}

The effective $Sp(1)$ connection -- $\tilde{q}^{A'}_{i B'}$ in local coordinates-- that appears in the extended \hk identity \eqref{ext-hk-id1}
will be given as
\begin{equation}
\begin{split}
&\tilde{q}^{A'}_{0 B'}=\frac{R}{V}\begin{pmatrix}  1 & 0  \\ 0 & 0\end{pmatrix},\\
&\tilde{q}^{A'}_{i B'}=0, \; i \neq 0.
\end{split} \label{Sp-conn-undef}
\end{equation}

The metric $g_{ij}$, holomorphic connection $\cA$, the map $e_{i AA'}$ and the $Sp(1)$ connection $\tilde{q}^{A'}_{i B'}$ completely specify the generalized hyperk\"ahler sigma model.

One can choose appropriate complex coordinates to directly show that the curvature $\F$ of the holomorphic connection $\cA$ is a $(1,1)$ form in the preferred complex structure characterized by the choice of $c^A$ and $c_A$. Alternatively, one may check that the $(2,0)$ part of $F$ vanishes.
\begin{equation}
\begin{split}
F^{ij} e_{i\;AA'} e_{j\; BB'} c^A c^B= F^{\gamma\sigma} e_{\gamma\; 2 A'} e_{\sigma\; 2 B'}+ F^{\sigma \gamma} e_{\sigma\; 2 A'} e_{\gamma\; 2 B'}=0, \forall A',B'.
\end{split}
\end{equation}

\subsubsection{$NW\circ \Omega \circ S$-deformed theory}

Now, let us rewrite the $NW\circ \Omega \circ S$-transformed theory as generalized \hk sigma model. We follow the same approach as outlined for the undeformed model -- namely, we read off the sigma model data from the action and supersymmetry transformation given in \eqref{NWOS-GH-bosaction}- \eqref{def-susy-ferm}.\\

Evidently, the bosonic action \eqref{NWOS-GH-bosaction} can be put in the form $$S_b= \frac{1}{8 \pi} \int_{\mathbb{R}^3} \de^3x \left[ g_{ij} \partial_{\mu} \varphi^i \partial^{\mu} \varphi^j +  \epsilon_{\mu\nu\rho} G^{\mu\nu} \partial^{\rho}\varphi^i \mathcal{A}_i\right]$$ and one can read off the metric $g$ on the target space $\ctM$ and the holomorphic connection $\cA$ on $\ctM$. In local coordinates, non-zero components of the metric $g$ are
\begin{equation}
\begin{split}
&g_{\gamma \gamma}=\frac{8 \pi R' V}{ \im \hat\tau}, \\
& g_{\sigma \sigma}=\frac{8 \pi R' V \hat\tau \bar{\hat\tau}}{ \im \hat\tau}-8 \pi R'(V-1) \im \hat{\tau} \sin^2{\theta}, \\
& g_{\gamma \sigma}=-\frac{4 \pi R' V (\hat\tau +\bar{\hat\tau})}{ \im \hat\tau},\\
& g_{u \sigma}=4 \pi R' (V-1)  \im \hat\tau \sin{2\theta},\\
&g_{uu}=8 \pi R' \im \hat\tau(V \sin^2{\theta}+\cos^2{\theta}),\\
& g_{vv}=-8 \pi R' \im \hat\tau.\\
\end{split} \label{NW-metric}
\end{equation}
while the non-vanishing components of the connection $\cA$ are
\begin{equation}
\begin{split}
&\mathcal{A}_{\sigma}=-8 \pi \I R'^2 (\gamma +\re \tau \tan{\theta} A_u),\\
& \mathcal{A}_{u}=-8 \pi \I R'^2 \tan{\theta} (\gamma+ \re \tau \tan{\theta} A_u).\\
\end{split}
\end{equation}
The non-vanishing components of the curvature $\F$ of $\cA$ are
\begin{equation}
\begin{split}
& \F_{\gamma \sigma}=- 8 \pi \I R'^2, \\
&\F_{\gamma u}=- 8 \pi \I R'^2 \tan{\theta}, \\
&\F_{\sigma u}=8 \pi \I R'^2 \re \tau \tan{\theta}.
\end{split}
\end{equation}

Let us define the following local coordinates on $\ctM$:  $z=R'((\gamma +\re \tau \tan{\theta} A_u) -\hat{\tau} (\sigma + \tan{\theta} A_u)), \bar{z}=R'((\gamma +\re \tau \tan{\theta} A_u) -\bar{\hat{\tau}} (\sigma + \tan{\theta} A_u))$. In the basis of local coordinates $\{z,\bar{z}, \varphi, \bar{\varphi}\}$, the non-zero components of the connection $\cA$ and the curvature $\F$ can be written as
\begin{empheq}[box=\fbox]{gather}
\mathcal{A}_z=-\mathcal{A}_{\bar{z}} =8 \pi \I \frac{(\hat{\tau} \bar{z} - \bar{\hat{\tau}} z)}{(\hat{\tau} -\bar{\hat{\tau}})^2}, \label{conn-2}\\ \nonumber
\F_{z\bar{z}} = -\frac{8 \pi \I}{(\hat{\tau} -\bar{\hat{\tau}})}.
\end{empheq}
The fermions $\psi^{E'}_{\alpha}, \bar{\psi}^{E'}_{\alpha}$ and the supersymmetry parameters $\zeta^E_\alpha, \bar{\zeta}^E_\alpha$ can be defined in precisely the same way as in the undeformed case, along with related antisymmetric pairings $\epsilon_{A'B'}$ and $\epsilon_{AB}$.\\

The map $e$ can again be read off from the supersymmetry rules \eqref{def-susy-bos} -- \eqref{def-susy-ferm}.
\begin{equation}
\begin{split}
&e^{AA'}_i \de \varphi^i=\begin{pmatrix} e^{11'}_u(V,\theta)\de A_u +e^{11'}_v(V,\theta) \de A_v &  \I e^{21'}_\gamma(V,\theta) \de \gamma +\I e^{21'}_\sigma(V,\theta) \de \sigma \\ \I e^{12'}_\gamma(V,\theta) \de \gamma +\I e^{12'}_\sigma(V,\theta) \de \sigma &  e^{22'}_u(V,\theta)\de A_u +e^{22'}_v(V,\theta) \de A_v\end{pmatrix}, \\
&e_{i\;AA'}\de \varphi^i=N\begin{pmatrix} -  e^{22'}_u(V,\theta)\de A_u - e^{22'}_v(V,\theta) \de A_v& \I e^{12'}_\gamma(V,\theta) \de \gamma +\I e^{12'}_\sigma(V,\theta) \de \sigma  \\  \I e^{21'}_\gamma(V,\theta) \de \gamma +\I e^{21'}_\sigma(V,\theta) \de \sigma &  -e^{11'}_u(V,\theta)\de A_u -e^{11'}_v(V,\theta) \de A_v \end{pmatrix},\\
&e^{AA'}_0=0,\; e_{0AA'}=0.
\end{split} \label{e-def-1}
\end{equation}
where the functions $e^{AB'}_i$ can be explicitly written as functions of the $\Omega$-deformation parameter $\theta$.
\begin{equation}
\begin{split}
&e^{11'}_u(V,\theta)=\sqrt{V}(1+\cos{\theta}(\frac{1}{\sqrt{V}} -1)),\\
& e^{11'}_v(V,\theta)=-1,\\
&e^{22'}_u(V,\theta)=V(1-\cos{\theta}(\frac{1}{\sqrt{V}} -1)),\\
& e^{22'}_v(V,\theta)=\sqrt{V},\\
&e^{12'}_\gamma(V,\theta)=\I V (\im \hat\tau)^{-1},\\
&e^{12'}_\sigma(V,\theta)=-\I V (\im \hat\tau)^{-1}(\hat\tau +\I \im \hat\tau (1-\cos{\theta})(\frac{1}{\sqrt{V}} -1)),\\
&e^{21'}_\gamma(V,\theta)=\I \sqrt{V} (\im \hat\tau)^{-1},\\
& e^{21'}_\sigma(V,\theta)=-\I \sqrt{V} (\im \hat\tau)^{-1}(\bar{\hat\tau} -\I \im \hat\tau (1+\cos{\theta})(\frac{1}{\sqrt{V}} -1)).
\end{split} \label{e-def-2}
\end{equation}
The fermions and SUSY parameters in the gauge theory are related to those in the sigma model in the following fashion:
\begin{equation}
\begin{split}
& \lambda^a_{\alpha} =\frac{1}{\sqrt{2V}} \left(\psi^{2'}_\alpha,\bar \psi_{\alpha}^ {2'} \right),\\
& \bar{\lambda}^a_{\alpha}=\frac{1}{\sqrt{2}} \left(\psi^{1'}_\alpha, \bar \psi_{\alpha}^ {1'} \right),\\
& \hat\zeta^a_{\alpha}= \frac{\sqrt{V}}{\sqrt{2}} \left(-\zeta^2_{\alpha}, -\bar{\zeta}^2_{\alpha}\right),\\
&\hat{\bar{\zeta}}^a_\alpha= \frac{\sqrt{V}}{\sqrt{2}} \left(\zeta^1_{\alpha}, \bar{\zeta}^1_{\alpha}\right).
\end{split}
\end{equation}

In the generalized sigma model, the constraint on the supersymmetry parameters $(\zeta^\alpha_A, \bar\zeta_\alpha^A)$ may be written as 
\begin{equation}
\begin{split}
&c^A \zeta^\alpha_A=0,\; c_A \bar\zeta_\alpha^A=0\\
&c^A=(-\sin{\frac{\theta}{2}},\cos{\frac{\theta}{2}}), \;c_A=(\cos{\frac{\theta}{2}},-\sin{\frac{\theta}{2}}).
\end{split}
\end{equation}

Since  $\zeta^a_{\alpha}$ is a constant spinor, the above identification implies that $\zeta^E_{\alpha}, \bar{\zeta}^E_{\alpha}$ must depend on $\varphi^0$ --  $\zeta^E_{\alpha}, \bar{\zeta}^E_{\alpha}$ obey equations of the form $\partial_{\mu} \zeta^E + f(\varphi^0) \partial_{\mu} \varphi^0 \zeta^E=0$, i.e.
\begin{equation}
\begin{split}
&\partial_{\mu} \zeta^E + \frac{1}{2} \frac{\partial_{\mu} V}{V}\zeta^E=0,\\
&\partial_{\mu} \bar{\zeta}^E + \frac{1}{2} \frac{\partial_{\mu} V}{V}\bar{\zeta}^E=0,\\
\implies & f(\varphi^0)=\frac{R'}{2V}.
\end{split}
\end{equation}

The effective $Sp(1)$ connection -- $\tilde{q}^{A'}_{i B'}$ in local coordinates-- that appears in the extended \hk identity \eqref{ext-hk-id1}
can be read off from the fermionic action
\begin{equation}
\begin{split}
&\tilde{q}^{A'}_{0 B'}=\frac{R'}{V}\begin{pmatrix}  1-\frac{(1-\cos{\theta})}{4} & -\frac{\sin{\theta}}{2\sqrt{V}}  \\ -\frac{\sin{\theta} \sqrt{V}}{2} & \frac{(1-\cos{\theta})}{4}\end{pmatrix},\\
&\tilde{q}^{A'}_{i B'}=0, \; i \neq 0.
\end{split}
\end{equation}
Note that the connection reduces to the form \eqref{Sp-conn-undef} in the limit $\theta \to 0$.\\

As before, the metric $g$, the surjection $e$, the connection $\cA$ and the $Sp(1)$ connection $\tilde{q}^{A'}_{i B'}$ completely specifies the sigma model data.

\subsubsection{Geometry of the constant $\varphi^0$ slice}

The target-space metric for the generalized \hk sigma model  a priori looks complicated -- in particular, it isn't clear if $\ctM[\varphi^0] \cong \R^2 \times T^2$, where $\ctM[\varphi^0]$ denotes the restriction to constant $\varphi^0$. The $\mathbb{R}^2 \times T^2$ structure of the fiber for constant $\varphi^0$ can be made manifest by another redefinition of fields in the following fashion:
\begin{equation}
\begin{split}
&\sigma \to \sigma -\frac{(V-1)\sin{2\theta}}{2(V\cos^2{\theta}+\sin^2{\theta})} A_u\\
&\gamma \to \gamma -\re \tau \frac{(V-1)\sin{2\theta}}{2(V\cos^2{\theta}+\sin^2{\theta})} A_u\\
&A_u \to \sqrt{\frac{(V\cos^2{\theta}+\sin^2{\theta})}{V}} A_u
\end{split}
\end{equation}

For the above choice of local coordinates, the target space metric assumes the following form:
\begin{equation}
\begin{split}
&g_{0u}=-\frac{8 \pi R' \im \hat\tau  \sin^2{\theta}}{2 V \left(\sin^2{\theta}+V \cos^2{\theta}\right)} A_u,\\
&g_{0\sigma}=-\frac{8 \pi R' \im \hat\tau \sin{2\theta}}{2\sqrt{V} \sqrt{V \cos^2{\theta}+\sin^2{\theta}}} A_u,\\
&g_{00}=\frac{8 \pi R' \im \hat\tau \sin^2{\theta} \left(4V\cos^2{\theta}+\sin^2{\theta}\right)}{4V^2 \left(V \cos^2{\theta} + \sin^2{\theta}\right)^2},\\
&g_{\gamma \gamma}=\frac{8 \pi R' V}{ \im \hat\tau}, \\
& g_{\sigma \sigma}=\frac{8 \pi R' V \hat\tau \bar{\hat\tau}}{ \im \hat\tau}-8 \pi R'(V-1) \im \hat{\tau} \sin^2{\theta}, \\
& g_{\gamma \sigma}=-\frac{4 \pi R' V (\hat\tau +\bar{\hat\tau})}{ \im \hat\tau},\\
&g_{uu}=8 \pi R' \im \hat\tau,\\
& g_{vv}=-8 \pi R' \im \hat\tau.
\end{split}
\end{equation}
The metric on the constant $\varphi^0$ slice clearly has the form of a standard metric on $\mathbb{R}^2 \times T^2$. The modular parameter $\tilde{\tau}$ of the torus $T^2$ can be computed from the metric components $g_{\gamma \gamma}, g_{\sigma \sigma}, g_{\gamma \sigma}$ given above.
\begin{empheq}[box=\fbox]{gather}
\re \tilde{\tau} =\re {\tau},\\ \nonumber
\im \tilde{\tau} =\im {\tau} \sqrt{\frac{(V\cos^2{\theta}+\sin^2{\theta})}{V \cos^2{\theta}}}.
\end{empheq}

\section{The NUT operator}
In the previous section, we have described in detail the local action of the generalized \hk sigma model that results from $NW\circ \Omega \circ S$- deformation of an $\N=2, U(1)$ SYM on Gibbons-Hawking space. Now, we will analyze the other essential ingredient of the sigma model path integral as mentioned in the introduction -- namely, the NUT operator $Q(\varphi(0))$.

As discussed in \cite{Dey:2014lja}, $Q(\varphi)$ may be determined explicitly by computing the partition function $\Psi$ of the deformed 
$U(1)$ gauge theory on a particular Gibbons-Hawking space, namely Taub-NUT space,
characterized by the harmonic function $V(\vec{x}) = 1 + R/r$.

The partition function $\Psi$ of the deformed theory may be computed directly from the UV description. For a $U(1)$ SYM, the theory receives no quantum corrections and $\Psi$ is a classical object receiving contributions from $U(1)$ instantons localized around the NUT center. The answer a priori depends on the choices of boundary conditions at spatial infinity:   $u$ and $v$ which give the asymptotic values of the
two real scalars of the theory; $\theta_e$ which gives the asymptotic value of the
holonomy of the $U(1)$ gauge field around the circle fiber, and $\theta_m$ which is associated with the dual photon. In addition, one expects the answer to depend on the angle $\theta$ parametrizing the $NW \circ \Omega \circ S$-deformation.

On the other hand, in the IR, the partition function only receives contribution from constant fields. The local action discussed above vanishes on
constant field configurations and the answer comes solely from the boundary term on the $S^2$ cut out
around the NUT center , i.e. the NUT operator.  \\

Since $U(1)$ SYM is a free theory, the UV and the IR answers for the partition function must agree and therefore we must have
\begin{equation}
\Psi = e^Q.
\end{equation}
which determines $Q$.

\subsection{UV computation for undeformed theory}
Let us consider the undeformed theory first. In this case, the solutions to the BPS equations in 3D which descend from 4D instantons localized around the NUT center are
\begin{equation}
\begin{split}
&\gamma=\frac{\theta_m -\tau \theta_e}{4 \pi R} -\frac{2\pi n \tau}{4 \pi R} +\frac{(\theta_e + 2n\pi)\tau}{4\pi RV},\\
&\sigma=\frac{(\theta_e + 2n\pi)}{4\pi RV},\\
&A_u =u,\\
&A_v=v.
\end{split} \label{undef}
\end{equation}
The classical action evaluated on this BPS configuration is
\begin{equation}
\begin{split}
S^{(n)}_{classical} &= -2\I R^2\int_{\mathbb{R}^3} \gamma \wedge \de\sigma \wedge dB =- 2\I R\int_{\mathbb{R}^3} \gamma \wedge \de\sigma \wedge \star_3 dV\\
&= \frac{\I}{2\pi} \left[(-\tau \theta_e^2/2+\theta_e \theta_m)+ 2\pi n (\theta_m - \tau \theta_e) -2\pi^2 n^2 \tau \right]
\end{split}
\end{equation}

Therefore, the partition function computed from the UV description is 
\begin{equation}
\begin{split}
\Psi\left(\theta_e,\theta_m,\tau \right)=&\sum_{n \in \mathbb{Z}} e^{-S^{(n)}_{classical}}\\
=&e^{\frac{\I}{2\pi}(\tau \theta_e^2/2-\theta_e \theta_m)}\sum_{n \in \mathbb{Z}} e^{\I \pi n^2 \tau - 2\pi \I n (2y)} \\
=& e^{8 \I \pi X(y,\bar{y},\tau)}  \Theta (\tau,2y )
\end{split}
\end{equation}

where we have defined
\begin{equation}
\begin{split}
&y=\frac{\theta_m-\tau \theta_e}{4\pi}, \\
&\bar{y}=\frac{\theta_m-\bar{\tau} \theta_e}{4\pi}\\
&X(y,\bar{y},\tau)=\frac{\tau y^2 -2\bar{\tau}y^2-\tau \bar{y}^2 + 2\bar{\tau} y \bar{y}}{2(\bar{\tau}-\tau)^2}
\end{split}
\end{equation}
and $\Theta (\tau,y ) = \sum_{n \in \mathbb{Z}} e^{\I \pi n^2 \tau - 2\pi \I n y}$ is the Jacobi Theta function. Following the discussion above, one concludes that
\begin{equation}
e^Q = e^{8 \I \pi X(y,\bar{y},\tau)}  \Theta (\tau,2y )
\end{equation}

One can readily check that $e^Q$ is a holomorphic section of a line bundle $\cL^{-1}$ where $\cA$ is a connection on the line bundle $\cL$.
\begin{equation}
(\partial_{\bar{y}}- \cA_{\bar{y}}) e^Q =0.
\end{equation}

\subsection{UV computation for deformed theory}
Next, we proceed to compute the NUT operator $e^{\widetilde{Q}}$ for the $NW\circ \Omega \circ S$-deformed theory -- the answer turns out to be a deformed version of the operator $e^Q$ computed above. The bosonic action of the theory dimensionally reduced to 3D is given in \eqref{NWOS-GH-bosaction}. Recall that this action can be obtained by a formal substitution of fields in the undeformed 3D action.
\begin{equation}
\begin{split}
&\sigma^{(undef)}=\cos{\theta} \sigma + \sin{\theta} A_u,\\
&A^{(undef)}_u=\cos{\theta}  A_u - \sin{\theta} \sigma, \\
&\gamma^{(undef)}=\cos{\theta} \gamma + \re \tau \sin {\theta} A_u.
\end{split} \label{formal - NW+Omega}
\end{equation}
while the deformed coupling and the deformed radius are given as
\begin{equation}
\begin{split}
& \hat{\tau}= \re \tau + \I \im \hat{\tau},\\
& \im \hat{\tau} = \im \tau \cos{\theta},\\
& R'= R\cos{\theta}.
\end{split}
\end{equation}

Inverting the equations \eqref{formal - NW+Omega}, we have
\begin{equation}
\begin{split}
&\sigma=\cos{\theta} \sigma^{(undef)} - \sin{\theta} A^{(undef)}_u\\
&A_u=\sin{\theta} \sigma^{(undef)} +\cos{\theta} A^{(undef)}_u \\
&\gamma=\frac{1}{\cos{\theta}}\left(\gamma^{(undef)} - \re \tau \sin^2{\theta} \sigma^{(undef)} - \re \tau \cos{\theta} \sin{\theta} A^{(undef)}_u\right)
\end{split}
\end{equation}

The solutions to the BPS equations for fields $\sigma, A_u, \gamma$ in the $NW\circ \Omega \circ S$-transformed theory can now be obtained from the corresponding solutions in the undeformed theory using the above equations and imposing appropriate boundary conditions. Therefore,
\begin{equation}
\begin{split}
&\sigma=\frac{\alpha}{4\pi R'V} + \frac{\alpha \sin^2{\theta}}{4\pi R'} (1-\frac{1}{V})- \tan{\theta} u,\\
&A_u= u + \frac{\alpha \sin{\theta}}{4\pi R} (-1+\frac{1}{V}),\\
&\gamma= \frac{\theta_m}{4\pi R'} + \frac{\alpha \hat{\tau}}{4\pi R'} (\frac{1}{V} -1) + \frac{\re \tau \sin^2{\theta} \alpha}{4\pi R'} (1-\frac{1}{V}).
\end{split}
\end{equation}
where $\alpha= (\theta_e + 2n\pi) + 4\pi R' u \tan{\theta}$. Note that the boundary conditions for the fields as $r \to \infty$ are $\sigma \to \frac{(\theta_e + 2n\pi)}{4\pi R'}$, $\gamma \to \frac{\theta_m}{4\pi R'}$ and $A_u \to u$.

Plugging in the above solution in the bosonic action of \eqref{NWOS-GH-bosaction}, one finds that the only contribution to the classical action for the above BPS solutions comes from the $S_{GH}$ (contributions from other terms add up to zero). Define
\begin{equation}
\begin{split}
& z=(\theta_m +4\pi R' u \re \tau \tan{\theta}) -\hat{\tau} (\theta_e+4\pi R' u \tan{\theta})=\widetilde{\theta}_m -\hat{\tau} \widetilde{\theta}_e, \\
&X(z,\bar{z},\hat{\tau})= \frac{\hat{\tau} z^2 -2\bar{\hat{\tau}} z^2-\hat{\tau} \bar{z}^2 + 2\bar{\hat{\tau}} z \bar{z}}{2(\bar{\hat{\tau}}-\hat{\tau})^2}.
\end{split}
\end{equation}

The classical action can therefore be written as
\begin{equation}
\begin{split}
S^{(n)}_{classical} =& \frac{\I}{2\pi} \left[(-\hat{\tau} \widetilde{\theta}_e^2/2+\widetilde{\theta}_e \widetilde{\theta}_m)+ 8\pi^2 n z -2\pi^2 n^2 \hat{\tau} \right]\\
=& 8\pi \I X(z,\bar{z},\hat{\tau})  + 4\pi \I n z - \I \pi n^2 \hat{\tau}
\end{split}
\end{equation}
Therefore, the NUT operator of the deformed theory will be given as
\begin{equation}
\begin{split}
e^{\widetilde{Q}}=\Psi\left(\theta_e,\theta_m,u,\hat{\tau}, \theta \right)=\sum_{n \in \mathbb{Z}} e^{-S^{(n)}_{classical}}=e^{8 \I \pi X(z,\bar{z},\hat{\tau}, \theta)}  \Theta (\hat{\tau},2z )
\end{split}
\end{equation}

One can now check that the NUT operator $e^{\widetilde{Q}}$ is a holomorphic section of a line bundle $\cL^{-1}$, where the 1-form $\cA$ deduced in \eqref{conn-2} is a connection on the line bundle $\cL$.
\begin{equation}
\left(\partial_{\bar{z}} - \mathcal{A}_{\bar{z}}\right) e^{\widetilde{Q}} =0
\end{equation}
where $\cA_z=-\cA_{\bar{z}} =8 \pi \I \frac{(\hat{\tau} \bar{z} - \bar{\hat{\tau}} z)}{(\hat{\tau} -\bar{\hat{\tau}})^2}$. The NUT operator for the deformed theory is therefore again a Jacobi Theta function but with arguments which explicitly depend on the $\Omega$-deformation parameter. As expected, the deformed operator reduces to the undeformed one in the limit $\theta \to 0$.

\appendix

\section{Dualizing a $U(1)$ theory on a Gibbons-Hawking space}\label{dualization}
The bosonic action of a $\N=2, U(1)$ gauge theory on a Gibbons-Hawking (GH) space is
\begin{equation}\label{bosact-GH}
S_{boson}= \frac{\im \tau}{4\pi}\int_{GH} \left[F^{(4)}\wedge \star^{(4)} F^{(4)} + \de A_u\wedge \star^{(4)} \de A_u -  \de A_v\wedge \star^{(4)} \de A_v\right] + \I \frac{\re\tau}{4\pi}\int_{GH} \left[F^{(4)}\wedge F^{(4)}\right].
\end{equation}
The metric on a Gibbons-Hawking space is
\begin{equation}
ds^2= V(\vec{x}) (\de \vec{x})^2 +R^2 V^{-1}(\vec{x}) \Theta^2,
\end{equation}
where $V(\vec{x}) \in \Omega^0(\mathbb{R}^3)$,$\Theta= \de \chi + B$, with $B \in \Omega^1(\mathbb{R}^3)$ and $\star^{(3)} \de B =\frac{1}{R} \de V$ -- implying that $V(\vec{x})$ is a harmonic function in $\R^3$. It is convenient to decompose the $p$-forms on GH in terms of forms on the $\R^3$ base and express four-dimensional star operator in terms of star operator on $\R^3$.

For a 1-form $\alpha \in \Omega^1(GH)$,
\begin{align}
&\alpha=\alpha^{(3)} + \alpha' \Theta\\
&\star^{(4)}(\alpha)= R\star^{(3)}(\alpha^{(3)})\wedge \Theta -V^2 R^{-1} \star^{(3)} \alpha'
\end{align}
where $\alpha^{(3)} \in \Omega^1(\mathbb{R}^3)$, $\alpha' \in \Omega^0(\mathbb{R}^3)$ and
$\star^{(3)}$ is the star operator for $\mathbb{R}^3$.\\

 For a 2-form $\beta \in \Omega^2(GH)$, one can show that
\begin{align}
&\beta=\beta^{(3)} + \beta'\wedge \Theta, \\
&\star^{(4)}(\beta) = R V^{-1}\star^{(3)}(\beta^{(3)})\wedge \Theta +V R^{-1} \star^{(3)} \beta',
\end{align}
where $\beta^{(3)}\in \Omega^2(\mathbb{R}^3)$ and $\beta' \in \Omega^1(\mathbb{R}^3)$.

 Decomposing $F^{(4)}$ as $F^{(4)}= F^{(3)} -R\;\de\sigma \wedge \Theta$, with $\sigma \in \Omega^0(\mathbb{R}^3)$, we have
\begin{align}
&\frac{1}{4\pi}\int_{S^1}\left[F^{(4)}\wedge \star^{(4)} F^{(4)}\right]=R \;V^{-1}F^{(3)}\wedge \star^{(3)} F^{(3)} +R V \; \de\sigma \wedge \star^{(3)}\de{\sigma},\\
&\frac{1}{4\pi}\int_{S^1}\left[F^{(4)}\wedge F^{(4)}\right]= -2 R \;F^{(3)} \wedge \de\sigma,\\
&\frac{1}{4\pi}\int_{S^1} \left[\de A_u\wedge \star^{(4)} \de A_u -  \de A_v\wedge \star^{(4)} \de A_v\right]=R (\de A_u\wedge \star^{(3)} \de A_u -  \de A_v\wedge \star^{(3)} \de A_v).
\end{align}
Collecting all terms, the bosonic action \eqref{bosact-GH}, dimensionally reduced to three dimensions is
\begin{equation}
\begin{split}
S_{boson}=&R\;\im \tau \int_{\mathbb{R}^3} \left[V^{-1}F^{(3)}\wedge \star^{(3)} F^{(3)} + V\de \sigma \wedge \star^{(3)}\de {\sigma}+\de A_u\wedge \star^{(3)} \de A_u -  \de A_v\wedge \star^{(3)} \de A_v\right]\\
&-\I R\;\re\tau\int_{\mathbb{R}^3} 2F^{(3)}\wedge \de \sigma
\end{split}
\end{equation}
To dualize the gauge field, we add the following term to the action 
\begin{equation}
\begin{split}
S_{dual}=&-2 \I R\int_{\mathbb{R}^3}(\gamma dF^{(3)}+R\;\gamma \wedge \de \sigma \wedge \de B)\\
=&2\I R\int_{\mathbb{R}^3}(\de\gamma \wedge F^{(3)}-R\gamma \wedge \de \sigma \wedge dB)-2\I R \int_{\infty} \gamma F^{(3)}.
\end{split}
\end{equation}
The scalar $\gamma$ appears as a Lagrange multiplier whose equation of motion imposes the Bianchi identity on $F^{(3)}$, namely
\begin{equation}
\de F^{(3)} + R \de \sigma \wedge \de B=0.
\end{equation}
One can now simply integrate out $F^{(3)}$ using its equation of motion in favor of the scalar $\gamma$. Modulo the boundary term, equation of motion for $F^{(3)}$ reads
\begin{equation}
F^{(3)}=-\I V (\im \tau)^{-1} \star(\de\gamma - (\re \tau) \de\sigma).
\end{equation}

Integrating out $F^{(3)}$ using the above equation of motion, the dualized 3D action reads:
\begin{equation}
\begin{split}
S_{boson}= &R \int_{\mathbb{R}^3} \left[V (\im \tau)^{-1} |\de\gamma - \tau \de\sigma|^2 + \de A_u\wedge \star^{(3)} \de A_u -  \de A_v\wedge \star^{(3)} \de A_v\right]\\
& -2 \I R^2\int_{\mathbb{R}^3} \gamma \wedge \de\sigma \wedge dB + S_{boundary},
\end{split}
\end{equation}
where
\begin{equation}
S_{boundary} = -2 \I R\int_{\infty} \gamma F^{(3)}.
\end{equation}

\subsubsection*{The $\theta_m$ term and boundary conditions at spatial infinity}

Consider adding to the 4D action the following boundary term :
\begin{equation}
\Delta S = \I \frac{\theta_m}{8\pi^2 R} \int_{{GH}_{\infty}} R \Theta \wedge F^{(4)} \label{bdry-NUT}
\end{equation}
where ${GH}_{\infty}$ denotes the $S^1$ bundle on $S^2$ at the $r \to \infty$ boundary of the Taub-NUT space.
Since $F^{(4)}= F^{(3)} - R \, \de\sigma \wedge \Theta$, we have
\begin{equation}
\Delta S= \I \frac{\theta_m}{8\pi^2 R} \int_{X_{\infty}} R\Theta \wedge F^{(3)} = \I \frac{\theta_m}{2\pi}  \int_{{\infty}} F^{(3)}.
\end{equation}
$\Delta S$ and $S_{boundary}$ will cancel each other if at $r \to \infty$, $\gamma$ and $\sigma$ have the boundary conditions
\begin{equation}
\boxed{\gamma \to \frac{\theta_m}{4\pi R}, \qquad \sigma \to \frac{\left(\theta_e+2n\pi\right)}{4\pi R}}
\end{equation}

Therefore, the final form of the dualized 3D action is
\begin{empheq}[box=\fbox]{gather}
S_{boson}= R \int_{\mathbb{R}^3} \left[V (\im \tau)^{-1} \lvert \de\gamma - \tau \de\sigma \rvert^2 + \de A_u\wedge \star^{(3)} \de A_u -  \de A_v\wedge \star^{(3)} \de A_v\right] \\
- 2 \I R^2\int_{\mathbb{R}^3} \gamma \wedge \de\sigma \wedge \de B.  \nonumber
\end{empheq}

The dualized 3D action for the $\Omega$-deformed theory (before any change of variables) is obtained from the above action by the formal substitution: $A_u \to A_u - \epsilon R \sigma$. Therefore,
\begin{empheq}[box=\fbox]{gather}
S^{\Omega}_{boson}= R \int_{\mathbb{R}^3} \left[\frac{V}{\im \tau} \lvert \de\gamma - \tau \de\sigma \rvert^2 + \de (A_u - \epsilon R \sigma)\wedge \star^{(3)} \de (A_u-\epsilon R \sigma) -  \de A_v\wedge \star^{(3)} \de A_v\right] \\
- 2 \I R^2\int_{\mathbb{R}^3} \gamma \wedge \de\sigma \wedge \de B. \nonumber
\end{empheq}

The scalars $\gamma$ and $\sigma$ are periodic scalars : $\sigma \sim \sigma + \frac{1}{2R}, \gamma \sim \gamma + \frac{1}{2R}$. The action is invariant under shift of $\sigma$, namely 
\begin{equation}
S(\sigma + 1/R) = S(\sigma). \label{sigma-period}
\end{equation}
However, under shift of $\gamma$ we have 
\begin{equation}
S(\gamma + 1/R)=S(\gamma) + k\theta_e + 2\pi n \label{gamma-period}
\end{equation}
where $k=\frac{1}{4\pi} \int \de B$. However, the total sigma model action \eqref{sigma-action-full} is still invariant under the shift since the NUT operator transforms in precisely the right way to cancel off the extra contributions in \eqref{gamma-period}.

\section{6D,4D and 3D Spinors}\label{6d4d3d}
In this section, we explain our conventions regarding 6D spinors and their dimensional reduction to 4D and 3D that will be useful in the main text.  We choose the following signature for the metric on a 6D flat space $\eta_{MN}=\left(-1,1,1,1,1,1\right)$ where $M,N=0,1,2,3,4,5$. The fermionic field $\psi_a$ is a symplectic Majorana-Weyl spinor which transforms as a doublet of the $SU(2)$ R-symmetry. 

\begin{equation}
\begin{split}
&\Gamma^7 \psi_a =\psi_a \; (\mbox{Weyl condition})\\
&\psi^{a,\; T}C^{-}_6 \epsilon_{ab}= (\psi^{b\;})^{\dagger} \I\Gamma^0 \equiv \bar{\psi}_b  \;(\mbox{Symplectic Majorana condition})
\end{split}
\end{equation}
where $\Gamma^7$ is the chirality matrix in 6D : $\Gamma^7=-\Gamma^0 \Gamma^1 \Gamma^2 \Gamma^3 \Gamma^4 \Gamma^5$. $C^{-}_6$ is the 6D charge conjugation which obeys $C^{-}_6 \Gamma^{M} (C^{-}_6)^{-1}=-(\Gamma^{M})^T$. \\

We choose the following representation of $\Gamma$ matrices and charge conjugation matrix in 6D:
\begin{equation}
\begin{split}
&\Gamma^M = \{ \I \sigma_2 \otimes \gamma^5,\;\sigma_1 \otimes \gamma^5,\;\textbf{1}_{2\times 2} \otimes \gamma^{m} \} \; (m=1,2,3,4;\; M=0,\ldots,5)\\
&\Gamma^7 = \sigma_3 \otimes \gamma^5,\; C^{-}_6 =\Gamma^0\Gamma^2\Gamma^4= \I \sigma_2 \otimes C^{-}_4,\\
&C^{-}_4= \gamma^1 \gamma^3\gamma^5,\\
&\gamma^5=-\gamma^1\gamma^2\gamma^3\gamma^4.
\end{split} \label{gamma6dto4d}
\end{equation}
where $\gamma^m$ are the 4D gamma matrices (Euclidean signature), $\gamma^5$ is the 4D chirality operator and $C^{-}_4$ is the 4D charge conjugation matrix. We choose the following convention for the gamma matrices and charge conjugation matrix in 4D:
\begin{equation}
\begin{split}
&\gamma^m= \left(\begin{matrix}
0 &\I \sigma^{m,\alpha\dot{\alpha}} \\
\I\bar{\sigma}^{m}_{\dot{\alpha}\alpha} & 0\\
\end{matrix}\right),\\
&\gamma_5=\left(\begin{matrix}
 \textbf{1}_{2\times 2} & 0\\
0 & -\textbf{1}_{2\times 2}\\
\end{matrix}\right),\\
&C^{-}_4= \left(\begin{matrix}
\epsilon_{\alpha\beta} & 0\\
0 & \epsilon^{\dot{\alpha} \dot{\beta}}\\
\end{matrix}\right),\\
&\sigma^m=(-\vec{\sigma}, \I),\; \bar{\sigma}^m=(\vec{\sigma}, \I),\; (m=1,2,3,4).
\end{split} \label{gamma4d}
\end{equation}
where $\vec{\sigma}$ are the Pauli matrices. The 4D spinor indices are raised and lowered by right action of the antisymmetric tensor $\epsilon_{\alpha\beta}$ (such that $\epsilon^{12}=\epsilon_{21}=1$) and $\epsilon_{\dot\alpha\dot\beta}$ (exactly the same convention). The above definition implies that 
$\sigma^m$ and $\bar{\sigma}^m$ are related in the following fashion:
\begin{equation}
(\bar{\sigma}^m)_{\dot\alpha\alpha}= \epsilon_{\beta\alpha}\epsilon_{\dot\beta\dot\alpha}(\sigma^m)^{\beta\dot\beta}
\end{equation}

For this representation, the symplectic Majorana-Weyl  spinor $\psi^a$ can be written as the doublet $\psi^a=\left(\begin{matrix}
\psi^a_1\\
\psi^a_2\\
\end{matrix}\right)$ and the Weyl condition implies that
\begin{equation}
\begin{split}
&\gamma^5  \psi^a_1=\psi^a_1 \implies \psi^a_1=\left(\begin{matrix}
\lambda^{a,\;\alpha}\\
0\\
\end{matrix}\right)\\
&\gamma^5  \psi^a_2=-\psi^a_2 \implies \psi^a_2=\left(\begin{matrix}
0\\
\bar{\lambda}_{\dot{\alpha}} ^{a}\\
\end{matrix}\right)\\
\end{split}\label{spinor6dto4d}
\end{equation}

The symplectic Majorana condition imposes reality conditions on $\lambda^{a,\;\alpha}$ and $\bar{\lambda}_{\dot{\alpha}} ^{a}$.
\begin{equation}
\begin{split}
&(\lambda^{\alpha, a})^{\dagger} = -\I \lambda^{\beta, b} \epsilon_{\beta \alpha} \epsilon_{ba}\\
&(\bar{\lambda}_{\dot{\alpha}}^{a})^{\dagger} =\I\bar{\lambda}_{\dot{\beta}}^{b} \epsilon_{ba} \epsilon^{\dot{\beta}\dot{\alpha}}\\
\end{split}\label{reality4d}
\end{equation}

If $(x_1,x_2,x_3,x_4)$ denote local coordinates on the manifold $M_4$, then let $x_2$ be the circle direction. The basic rules of dimensional reduction for the fermionic fields $\lambda^a$ and the gamma-matrices to 3D can be summarized as follows:
\begin{equation}
\begin{split}
&({\sigma}^m)^{\alpha\dot{\beta}} \to (\gamma^i)^{\alpha\beta}=(\gamma^i)^{\beta\alpha}=(-\sigma_1,-\sigma_3,\I)\; (m \neq 2) \\
&(\bar{\sigma}^m)_{\dot{\alpha}{\beta}} \to  (\gamma^i)_{\alpha\beta}=(\gamma^i)_{\beta\alpha}=(\sigma_1,\sigma_3,\I)\;(m \neq 2)\\
&({\sigma}^2)^{\alpha\dot{\beta}} =-\sigma_2 \to  \I \epsilon^{\alpha\beta},\; (\bar{\sigma}^2)_{\dot{\alpha}{\beta}}=\sigma_2 \to \I\epsilon_{\alpha\beta}\\
&\lambda^{a,\alpha} \to  \lambda^{a,\alpha}, \;\bar{\lambda}^{a}_{\dot{\alpha}} \to \bar{\lambda}^{a}_{{\alpha}}
\end{split}
\end{equation}
where we normalize the antisymmetric tensor $\epsilon_{\alpha\beta}$ as $\epsilon^{12}=\epsilon_{21}=1$.
The matrices $\gamma_i :=(\gamma_i)_{\alpha}^{\;\;\beta}=(-\sigma_3,\sigma_1,-\sigma_2)$ obey the following identities
\begin{equation}
\begin{split}
&\{\gamma_i,\gamma_j\}=\delta_{ij} \textbf{1} \;(\mbox{Clifford algebra})\\
&\gamma_i \gamma_j=\delta_{ij} \textbf{1} + i \epsilon_{ijk}\gamma_k \; (\mbox{Product Identity})
\end{split}
\end{equation}

\section{Killing Spinor on Omega-deformed NUT Space} \label{KS-omega}
Consider the six-dimensional space of the form $\mathcal{M}=TN \times S^1 \times S^1$, where the NUT space (TN) is non-trivially fibered over the first $S^1$ with the following fiber bundle metric,
\begin{equation}
ds^2_6=\sum^3_{m=0} (e^m_{(TN)} -\varepsilon V^m dx^4 )^2 + (du)^2 - (dv)^2
\end{equation}
If we choose $V$ to be the $U(1)$ isometry of the NUT space which leaves the Killing spinor on the NUT space invariant, i.e. $\mathcal{L}_V \eta_{Killing}=0$, then the components of $V=\frac{\partial}{\partial \chi}$ are,
\begin{equation}
V^0=0, V^1=0, V^2=0, V^3=\frac{R}{\sqrt{V}}
\end{equation}

We choose the orthogonal basis of 1-forms as follows:
\begin{align}
&e^0=e^0_{TN}=\sqrt{V} dr\\
&e^1=e^1_{TN}=r\sqrt{V} d\theta\\
&e^2=e^2_{TN} =r\sqrt{V}\sin{\theta} d\phi\\
&e^3=e^3_{TN}-\frac{\epsilon R}{\sqrt{V}}du=\frac{R}{\sqrt{V}}(d\chi - \cos{\theta} d\phi)-\frac{\epsilon R}{\sqrt{V}}du\\
&e^4=du\\
&e^5=dv
\end{align}

The independent, nonzero spin connections on undeformed NUT space are 
\begin{align}
&\omega^{10}=\frac{1}{rV} \frac{\de(r\sqrt{V})}{\de r} e^1=(1-\frac{R}{2rV})\de\theta\\
&\omega^{20}=\frac{1}{rV} \frac{\de(r\sqrt{V})}{\de r} e^2=(1-\frac{R}{2rV})\sin{\theta} \de\phi\\
&\omega^{30}=\frac{-1}{2V^{3/2}} \frac{\de V}{\de r} e^3=\frac{R^2}{2r^2V^2}(\de\chi - \cos{\theta} \de\phi)\\
&\omega^{13}=\frac{-R}{2r^2V^{3/2}} e^2=\frac{-R}{2rV}\sin{\theta} \de\phi\\
&\omega^{23}=\frac{R}{2r^2V^{3/2}} e^1=\frac{R}{2rV}\de\theta\\
&\omega^{12}=-\frac{R}{2r^2V^{3/2}} e^3 - \frac{\cos{\theta}}{r\sqrt{V}\sin{\theta}}e^2=-\frac{R^2}{2r^2V^2}\de\chi -(1-\frac{R^2}{2r^2V^2})\cos{\theta} \de\phi
\end{align}

The spin-connections of the $\Omega$-deformed space can be conveniently expressed in terms of the spin-connections on the NUT space in the following fashion:
\begin{align}
&\omega^{ab}= \omega^{ab}_{(TN)} -\frac{\epsilon R}{\sqrt{V}} \omega^{a3}_{(TN) b} e^4\\
&\omega^{30}=\omega^{30}_{(TN)} -\frac{\epsilon R^2}{2r^2V^2} e^4,\\
&\omega^{31}= \omega^{31}_{(TN)},\\
&\omega^{32}= \omega^{32}_{(TN)}, \\
&\omega^{\alpha m}=0\; (\alpha=4,5;\; m=0,1,2,3), \\
&\omega^{\alpha\beta}=0\;(\alpha,\beta=4,5).
\end{align}

The Killing spinor equation for the manifold is given as $\partial_{\mu}\eta +\frac{1}{4}\omega^{MN}_{\mu} \Gamma_{MN} \eta=0$. The components $\mu=r, \theta, \phi, \chi$ of this equation will be identical to that of the undeformed NUT space:
\begin{equation}
D_{\mu} \eta \equiv \partial_\mu \eta + \frac{1}{4}\omega^{MN}_{\mu} \Gamma_{MN} \eta=\partial_\mu \eta + \frac{1}{4}\omega^{mn}_{(TN)\mu} \Gamma_{mn} \eta=0 \label{NUTKS}
\end{equation}
The $u, v$ components are respectively,
\begin{eqnarray}
D_{u} \eta \equiv \partial_u \eta + \frac{1}{4}\omega^{MN}_{u} \Gamma_{MN} \eta=\partial_u \eta +\frac{\epsilon R^2}{4r^2V^2}(\Gamma_{12}- \Gamma_{30})\eta=0\\
D_{v} \eta \equiv \partial_v \eta=0
\end{eqnarray}
From the explicit expressions for spin connections, one can easily check that the combination 
$$D_u + \epsilon R D_y= \left(\partial_u + \frac{\epsilon R^2}{4r^2 V^2}(\Gamma_{12}-\Gamma_{30})\right)+\epsilon R\left(\partial_y +\frac{\epsilon R^2}{4r^2 V^2} (\Gamma_{30} - \Gamma_{12})\right)=\partial_u+\epsilon R \partial_y.  $$
We use this fact to simplify the $\Omega$-deformed fermionic action in the main paper.

The conditions for equations (\ref{NUTKS}) to have a solution, as we saw in the case undeformed NUT case, are:
\begin{equation}
(\Gamma_{12}-\Gamma_{30})\eta=0, \; (\Gamma_{01}-\Gamma_{32})\eta=0, \; (\Gamma_{02}-\Gamma_{13})\eta=0
\end{equation}
The condition for the Killing spinor to be independent of $u$ is given by the first of the above equations. Therefore, the condition for the existence of a Killing spinor on the six-dimensional manifold $\mathcal{M}$ is 
\begin{equation}
\Gamma_{0123}\eta=\eta
\end{equation}
which is identical to the condition for the existence of a Killing spinor in the undeformed case (i.e. for $\epsilon=0$). The Killing spinor can be explicitly written in terms of the local coordinates as
\begin{equation}
\eta= e^{-\theta \Gamma_{10}} e^{\phi \Gamma_{12}} \eta_0 \;\;\;\mbox{where} \;\;\Gamma_{0123}\eta_0=\eta_0
\end{equation}
To see what the above constraint $\Gamma_{0123}\eta_0=\eta_0$ means in terms of the 6D spinors defined before, note the identificaion $\Gamma_0 \to \Gamma^2$, $\Gamma_1 \to \Gamma^5$, $\Gamma_2 \to \Gamma^4$,$\Gamma_3 \to \Gamma^3$, where the $\Gamma^i$ are the gamma matrices defined in  Appendix \ref{6d4d3d}. The constraint therefore is $\Gamma^2\Gamma^5\Gamma^4\Gamma^3 \eta_0=\eta_0$ which is equivalent to $-\Gamma^2\Gamma^3\Gamma^4\Gamma^5 \eta_0=\eta_0$ or $\textbf{1}_{2\times 2} \otimes \gamma^{5} \eta_0=\eta_0$. Therefore, the 6D spinor satisfying the constraint is of the form,
\begin{equation}
\eta_0=\left(\begin{matrix}
\zeta\\
0\\
\end{matrix}\right)
\end{equation}

From a 4D perspective, the projection condition simply reduces to $\gamma^{5} \zeta =\zeta$ which restricts the supersymmetry to half.

\printbibliography

\end{document}